\documentclass[12pt]{iopart}

\usepackage{iopams}  
\usepackage{xypic}
\usepackage[matrix,frame,arrow]{xy}
\usepackage{iopams}

\newcommand{\qw}[1][-1]{\ar @{-} [0,#1]}

\newcommand{\gate}[1]{*{\xy *+<.6em>{#1};p\save+LU;+RU **\dir{-}\restore\save+RU;+RD **\dir{-}\restore\save+RD;+LD **\dir{-}\restore\POS+LD;+LU **\dir{-}\endxy} \qw}

\newcommand{\multigate}[2]{*+<1em,.9em>{\hphantom{#2}} \qw \POS[0,0].[#1,0];p !C *{#2},p \save+LU;+RU **\dir{-}\restore\save+RU;+RD **\dir{-}\restore\save+RD;+LD **\dir{-}\restore\save+LD;+LU **\dir{-}\restore}
\newcommand{\ghost}[1]{*+<1em,.9em>{\hphantom{#1}} \qw}

\newcommand{\Qcircuit}[1][0em]{\xymatrix @*[o] @*=<#1>}  
 \renewcommand{\Qcircuit}[1][0em]{\xymatrix @*=<#1>}

\newcommand{\pureghost}[1]{*+<1em,.9em>{\hphantom{#1}}}
\newcommand{\multiprepareC}[2]{*+<1em,.9em>{\hphantom{#2}}\save[0,0].[#1,0];p\save !C
  *{#2},p+RU+<0em,0em>;+LU+<+.8em,0em> **\dir{-}\restore\save +RD;+RU **\dir{-}\restore\save
  +RD;+LD+<.8em,0em> **\dir{-} \restore\save +LD+<0em,.8em>;+LU-<0em,.8em> **\dir{-} \restore \POS
  !UL*!UL{\cir<.9em>{u_r}};!DL*!DL{\cir<.9em>{l_u}}\restore}

\newcommand{\poloFantasmaCn}[1]{{{}^{#1}_{\phantom{#1}}}}

\def\Lin{\mathsf{Lin}}
\def\spc#1{\mathscr{#1}}
\def\Irr{\mathsf{Irr}}
\def\comment#1{}
\def\labell#1{\label{#1}}

\def\set#1{{\sf #1}}
\def\map#1{\mathcal #1} 
\def\grp#1{{\mathbf #1}}
\def\Cmplx{\ensuremath{\mathbb C}}

\def\Tr{\ensuremath{\mathrm{Tr}}}

\def\spc#1{\mathcal{#1}}
\def\<{\ensuremath{\langle}}
\def\>{\ensuremath{\rangle}}

\def\Span{\mathsf{Span}}

\def\Supp{{\mathsf{Supp}}}

\newtheorem{Example}{Example}
\newtheorem{Lemma}{Lemma}
\newtheorem{Theo}{Theorem}
\newtheorem{Corollary}{Corollary}
\newtheorem{Proposition}{Proposition}
\newtheorem{Def}{Definition}

\def\Proof{\medskip\par\noindent{\bf Proof. }}
\def\qed{$\,\blacksquare$\par}

\begin{document}

\title{Identification of a reversible quantum gate: assessing the resources}
\author{Giulio Chiribella} 
\address{Center for Quantum Information, Institute for Interdisciplinary Information Sciences, Tsinghua University, Beijing 100084, China. }
\author{Giacomo Mauro D'Ariano} 
\address{QUIT group, Dipartimento di Fisica ``A. Volta'', INFN Sezione di Pavia, via Bassi 6, 27100 Pavia, Italy.}  
 \author{Martin Roetteler}
\address{Microsoft Research, One Microsoft Way, Redmond, WA 98052, USA.}

\begin{abstract}
We assess the resources  needed to identify a reversible quantum gate among a finite set of alternatives, including in our analysis both deterministic and probabilistic  strategies.   
Among the probabilistic strategies, we consider  unambiguous gate discrimination---where errors are not tolerated but inconclusive outcomes are allowed---and  we prove that parallel strategies are sufficient to unambiguously identify the unknown gate with  minimum number of queries. This result is used to provide upper and lower bounds on the query complexity and on the minimum ancilla dimension.   
In addition, we introduce the  notion of generalized $t$-designs, which includes unitary $t$-designs and group representations as special cases.   For gates forming a generalized $t$-design we give an explicit expression for the maximum probability of correct gate identification and we  prove  that there is no gap between the performances of deterministic strategies an those of probabilistic strategies. 
 Hence, evaluating of the query complexity of perfect deterministic discrimination is reduced to  the easier problem of evaluating the query complexity of unambiguous discrimination.  Finally,  we consider discrimination strategies where the use of ancillas is forbidden, providing upper bounds on the number of  additional queries needed to make up for the lack of entanglement with the ancillas.

 \end{abstract}

\maketitle

\section{Introduction}

Identifying an unknown unitary evolution, available as a black box, is a fundamental problem in quantum theory \cite{Acin,EntImproves,bergouhill1,chefles1,EntNotNeeded,bergouhill2,chefles2,memorydisc,perfectduan}, with a wide range of applications in quantum information and computation.  
In quantum computation, 
the problem is known as  oracle identification  \cite{amb1,vandam,fahri,amb2,amb3} and is the core of paradigmatic quantum algorithms such as Grover's \cite{grover} and Bernstein-Vazirani's \cite{bv}. 
In quantum information processing, the identification of an unknown unitary gate plays a key role in the stabilizer formalism of quantum error correction \cite{stab1,stab2} and in its generalization to unitary error bases \cite{nice1,nice2,nice3,nice4,nice5}, in the security analysis of quantum cryptographic protocols that encode secret data into unitary gates \cite{bf,deng1,deng2,luca,clonunit,pira}, in the alignment of reference frames via quantum communication \cite{kempe,refframe,bagan,gloves,colorcod}, in  the design of quantum communication protocols that work in the absence of shared reference frames \cite{commun,speakable,skoti} and quantum machines that learn to execute a desired operation  from a training set of examples \cite{unitlearn}.  For all these applications, the crucial step is to find efficient strategies that discriminate among a set of unknown gates with minimum expenditure of resources. Typical resources considered are: the number of  black box queries needed to identify the unknown gate,  the number of time steps and the size of the auxiliary systems (ancillas) employed in the discrimination strategy, and the total number of elementary gates needed to implement the discrimination strategy.

A striking feature of gate discrimination is that any two distinct unitaries  can be perfectly distinguished from one another in a finite number of queries, either using  entanglement  \cite{Acin, EntImproves} or using a sequential strategy where different queries are called at different time steps \cite{EntNotNeeded}.   This feature implies that an unknown gate in a finite set  $\set U  :=  (U_x)_{x\in\set X}$ can be perfectly identified in a finite number of queries, e.g. by running  $|\set U|-1$ pairwise tests each of which eliminates one wrong alternative \cite{Acin}.  However, in terms of efficiency the method of pairwise elimination leaves  large room for improvement:   
For example, when the unitaries    are mutually orthogonal, one can identify the black box in a single query using an ancilla, following the lines of   the dense coding protocol \cite{dense}.  In general, finding the minimum number of queries needed for perfect discrimination is a hard problem:  solving it would automatically give a general solution for the query complexity of oracle identification.    
  One way to approach the problem is to consider the less demanding task of \emph{unambiguous gate discrimination} \cite{bergouhill1,chefles1,bergouhill2,chefles2,wang}, where the unknown gate is identified without errors but one allows for an inconclusive result.    General conditions  for unambiguous discrimination were given in Refs. \cite{chefles1,chefles2,wang} under the assumption that  the available queries are used in parallel. However, the case of general strategies and the quantification of the resources required for unambiguous gate discrimination have remained largely unaddressed up to now.  

In this paper we  provide a systematic study of the resources needed to identify an unknown gate, focussing in particular on the following resources:  number of queries, size of the ancillary systems, and number of time steps in the discrimination strategy.  We start from the observation that parallel strategies are sufficient for  unambiguous gate discrimination: if unambiguous discrimination can be achieved in $N$ queries, then it can be achieved by applying the $N$ queries in parallel (in general, using ancillas).   Based on the reductions to parallel strategies, we provide lower and upper bounds on the query complexity of unambiguous discrimination and on the size of the ancilla systems needed by the discrimination strategy.   The bounds are general and can often be improved in specific cases. Nevertheless, they  suffice to show that unambiguous discrimination of the gates $\set U $  is always possible with no more than $|\set U|-1$ queries. Since  $|\set U|-1$ is the minimum number of queries that would be needed by the method of pairwise elimination, our result shows that a joint discrimination strategy typically offers an advantage. 

After having discussed the resources for unambiguous discrimination, we ask under which conditions one can achieve the stronger task of perfect discrimination, where inconclusive outcomes are not allowed.  This is important because in practice the usefulness of unambiguous discrimination can be undermined by the fact that the inconclusive outcome occurs too frequently.  To this purpose, we introduce the notion of \emph{generalized $t$-designs}, which  includes as special cases the unitary $t$-designs of Refs.  \cite{tdesign1,tdesign2,tdesign3,tdesign4} and  all the examples where the unknown gates form a group \cite{covlik,entest,degiorgi,haya}.   Relative to  gate identification, generalized $t$-designs have three important features:  
 \begin{enumerate}
 \item there is no difference between the performances of deterministic and probabilistic strategies allowing for inconclusive outcomes
 \item there is no difference between the performances of strategies using the queries in parallel and general strategies using the queries in a sequence of time steps
 \item  there is a simple analytic formula for the maximum probability of correct gate identification with given number of queries. 
 \end{enumerate}
The feature \emph{i)}  implies that, if unambiguous discrimination is possible in $N$ queries, then also perfect deterministic discrimination must be possible in $N$ queries.  This result reduces the query complexity of  perfect discrimination to the query complexity of unambiguous discrimination, which is much simpler to evaluate.   The reduction to unambiguous discrimination has a fairly large range of applications, especially in the case when the set of gates  forms a group. 
  Particular examples are the group of all  Boolean oracles \cite{amb1}, the groups of linear \cite{bv} and quadratic \cite{quadraticmartin} Boolean functions, the group of permutations \cite{kempe}, and the group of all oracles corresponding to functions from a given finite set to another \cite{chefles2}.  
The feature  \emph{ii)}  implies that the number of time steps needed to  identify  a gate picked from a generalized $t$-design is minimum:  applying the queries in parallel one can reduce the discrimination strategy to three steps: the preparation of an entangled state, the parallel application of the unitary gates, and the execution of a suitable measurement.  Note that the number of time steps in a discrimination strategy should not be confused with the number of elementary gates needed to implement the strategy: preparing the joint state  and performing the joint measurement may require a large number of elementary gates. Nevertheless, the fact that in principle the number of time steps can be reduced to the minimum is an interesting and non-trivial property. In general, such a  property does not hold  when the unitaries do not form a generalized $t$-design: for example, using the available queries in parallel would spoil the quadratic speed-up in Grover's algorithm  \cite{brassard,zalka}. 
 
Finally, we address the quantification of resources for strategies where the use of ancillary systems is forbidden.  
   This analysis  is important for applications to reference frame alignment \cite{kempe,refframe,bagan,gloves,colorcod,qcmc} and quantum communication in the absence of  shared reference frames \cite{commun,speakable,skoti}. 
   Our contribution to these research topics is \emph{a)} to show that every gate discrimination using ancillas can be converted into a strategy using a number of extra queries to the unknown gate and \emph{b)} to provide bounds on the number of extra queries.   When the dimension of the ancilla used in the original strategy is large, we show  that the number of extra queries scales logarithmically with the ancilla dimension: a strategy using $N_A  \gg1$ ancillary qubits can be replaced by a strategy using $O(N_A)$ extra queries.   More specific bounds ben be obtained when the unitaries form a generalized $t$-design or a group.   In all these cases, we show that, again, there is no difference between the performances of deterministic strategies and those of probabilistic strategies allowing for inconclusive outcomes.

The paper is structured as follows. In section \ref{sec:results} we give a synopsis of the main results.  The basic facts about general gate discrimination strategies, along with the ovservation that unambiguous discrimination can be parallelized, are provided in Section \ref{sec:abstention}, and exploited  in Section \ref{sec:complexity} to derive upper and lower bounds on the query complexity of unambiguous discrimination.  In Section \ref{sec:designs} we introduce the notion of generalized $t$-designs, giving an explicit formula for the maximum probability of correct gate identification and showing that  parallel deterministic strategies achieve the same performances of arbitrary probabilistic strategies.  Bounds on the size of the ancilla needed for unambiguous/perfect discrimination with minimum queries are provided in Section \ref{sec:ancilla}, while Section \ref{sec:noancilla} considers discrimination schemes where the ancilla is not allowed, providing estimates for the query overhead.    
The conclusions are drawn in Section \ref{sec:conclusions}.  The Appendix contains all the technical proofs of the results presented in the paper.

\section{Main results}\label{sec:results}
We provide here a synopsis of the main results of the paper.  A more extended discussion, including the precise definition of the framework, additional results and applications will be the object of the following Sections. 

\medskip 

\noindent {\em Unambiguous gate discrimination: parallelizability and bounds on the query complexity.}  
We start by showing that unambiguous gate discrimination can be parallelized: if the gates in a given set can be distinguished unambiguously with $N$ queries, then they can be distinguished unambiguously by applying the queries in parallel, possibly using ancillas.  This fact is extremely useful to provide bounds of the query complexity of  unambiguous discrimination.   
Denoting by $N_{\min}$  the  minimum number of queries needed to unambiguously identify a unitary gate in the set   $\set U$, we prove the bounds 
\begin{equation}\label{ippodue}
|{\mathsf U}| \le  
\left(   \begin{array}{c}
  N_{\min} + d^2- 1 \\
 d^2-1 
 \end{array}
 \right)  
\end{equation}
and \begin{eqnarray}\label{unamb}  
N_{\min}  \le    |\set U|  -  \dim (\set U)  +1 ,      
\end{eqnarray}
where $d$ is the dimension of the Hilbert space where the gates act and $\dim (\set U)$ is the number of linearly independent unitaries in $\set U$.  Both bounds are tight, in the sense that for every size $|\set U|$ one can find a set of gates achieving the equality. 
The upper bound of Eq. (\ref{unamb}) proves that  a joint discrimination strategy typically needs less queries than a strategy based on pairwise eliminations. The bound of Eq. (\ref{ippodue}) contains implicitly a lower bound on $N_{\min}$,  which can be estimated as  
\begin{equation}
N_{\min}  > |  \set U|^{\frac1{ d^2 -1}}  -1   
\end{equation} 
When $d$ is fixed and $|\set U|$ is large, this estimate gives the actual scaling of the tight bound of Eq. (\ref{ippodue}).

In addition to the above bounds, we also provide a bound  in term of the maximum fidelity between pairs of gates.  
The bound is obtained from a simple observation about unambiguous state discrimination of pure states, which to the best of our knowledge  did not appear  in the previous literature on  the subject: 
the states in a generic set  $\{|\psi_x\> \}_{x\in\set X}$ can  be unambiguously discriminated using $N$ identical copies whenever $N $ satisfies
\begin{eqnarray}  
N  >  \frac{ \log  (|\set X|-1)  }{\log \left( F^{- \frac 12 }\right) }    \qquad  F : =  \max_{ x\not = y}  |\< \psi_x|  \psi_y\>|^2.
\end{eqnarray}  

In the case of gate discrimination,  this result can be applied to the set of bipartite states $ |\Psi_x\>  :=     (U_x\otimes I)  |\Psi\>$, where $|\Psi\>  $ is a bipartite input state.   
Optimizing over $|\Psi\>$, we then get the \emph{fidelity bound}
\begin{eqnarray}\label{fbound}
N_{\min}   \le     \left \lfloor   \frac{ \log ( |\set U|-1)  }{  \log  \left( F_{\set U}^{-1/2}\right)} \right \rfloor +1,  
\end{eqnarray}
where $F_{\set U}$ is the \emph{minimax fidelity}
$F_{\set U}  :  =\min_{  |\Psi\> \in  \spc H  \otimes \spc H , |\!|  \Psi |\!|  = 1} \max_{ x\not =y}     \left |  \<\Psi|    (  U_x^\dag U_y   \otimes I)  |\Psi\> \right|^2$.  
The fidelity bound is important because it connects a measure of pairwise distinguishability  with the the performances of general joint strategies for unambiguous discrimination. Moreover, in several examples it gives a better estimate than the linear bound of Eq. (\ref{unamb}).


 \medskip

\noindent {\em Generalized $t$-designs: maximum error probability and optimality of parallel deterministic strategies.} 
We introduce the notion of generalized $t$-designs, which  includes as special cases the unitary $t$-designs of Refs.  \cite{tdesign1,tdesign2,tdesign3,tdesign4} and  all the examples where the unknown gates form a group \cite{covlik,entest,degiorgi,haya}.  
When the unitary gates form a generalized $t$-design, we consider the problem of gate discrimination with minimum error probability, or, equivalently, with maximum probability of correct gate identification.  Optimizing over all possible discrimination strategies, we show that the  that maximum probability of correct identification with $N\le t$ queries (conditional to the occurrence of conclusive outcomes) is given by  
\begin{eqnarray}
p^{\max}_N  =  \frac{  \dim  \set   U_N}{|\set U|}  \qquad \set U_N  :  = \left (  U_x^{\otimes N}\right)_{x \in \set X}.
\end{eqnarray}
Moreover, we show that this optimum value can be achieved by a deterministic strategy that uses the $N$ queries in parallel.
As a consequence, this shows that \emph{i)} there is no difference between the performances of deterministic and probabilistic discrimination strategies, and \emph{ii)} there is no difference between the performances of parallel strategies and those of strategies using a sequence of multiple time steps.    
In particular, if a set of gates $\set U$ is  a generalized $|\set U|$-design, then there is no difference between perfect and unambiguous discrimination: whenever unambiguous discrimination is possible,  the probability of the inconclusive result can be reduced to zero. This result is important from the practical point of view, because unambiguous discrimination by itself may not be a useful primitive if the probability of the inconclusive result is too high. 
Moreover, thanks to the reduction to deterministic strategies, Eqs. (\ref{ippodue}), (\ref{unamb}) and (\ref{fbound}) become  bounds on the query complexity of  perfect deterministic gate discrimination. 
 \medskip 

\noindent {\em Minimum ancilla dimension.}  Another important resource, in addition to the number of  queries,  is the dimension of the ancilla needed to achieve gate discrimination \cite{entest,chen}. The ancilla dimension quantifies the extra memory space  used for the discrimination task.    We show that, when $N$ queries to the black boxes are used, the minimum ancilla dimension  can be upper bounded as
\begin{equation}\label{ancilla1}
d^{\min}_{A,N} \le  \left(  \begin{array}{c}   N  +  d - 1     \\ d-1 \end{array} \right).             
\end{equation}
Since the binomial can be upper bounded as $(N+1)^{d-1}$, our result implies that  the size of the ancilla scales at most polynomially in the number of queries. In other words, this means that the number of ancillary qubits needed for gate discrimination in $N$ queries is at most logarithmic in $N$.   
 The bound  is independent of  the gate set $\set U$.  When more information on the gates is available, further estimates can be provided:  For example, if the set $\set U$ is contained in a representation of a finite group $\set G$, the dimension of the ancilla can be upper bounded as 
 \begin{eqnarray}\label{ancilla2}
 d^{\min}_{A,N}  \le  \sqrt{  |\set G|},
 \end{eqnarray}
 independently of the number of queries and of the dimension of the Hilbert space.  This bound  provides a fast estimate of the ancilla dimension, and, in several situations, the estimate is actually accurate. For example, in the case of the Pauli matrices $\{I,X,Y,Z\}$ the bound gives correctly $d^{\min}_{A,N}  \le  2$, meaning that unambiguous discrimination is possible using a single ancilla qubit.  This is indeed what is achieved by  the dense coding protocol \cite{dense}.    An even stronger result holds if the unitaries in $\set U$ commute: in this case, no ancilla at all is needed, a result that was already known for discrimination strategies using parallel queries \cite{entest,chefles2}.   
 
\medskip  
\noindent {\em Ancilla-free gate discrimination.}  For applications in reference frame alignment \cite{kempe,refframe,bagan,gloves,colorcod,qcmc} and quantum communication in the absence of  shared reference frames \cite{commun,speakable,skoti}, it is useful to consider discrimination strategies that do not use ancillas, here referred to as \emph{ancilla-free}. 
In this case, one can make up for the lack of ancillas using a number of additional queries to the black box, using the invariant encoding of Ref. \cite{prlcommun}.    
For a strategy using a large number of ancilla qubits $N_A$, we show that the scaling of the minimum overhead  $\Delta N_{\min}$  is upper bounded as 
\begin{eqnarray}\label{O}
\Delta N_{\min}  \le O(N_A).
\end{eqnarray} 
In other words, the $N_A$ ancilla qubits can be replaced with (order of) $N_A$ extra queries to the black box, showing that the use of extra queries is a more powerful resource than the use of ancillas.   
In addition, we provide the ancilla-free version of the upper bounds of Eqs. (\ref{unamb}) and (\ref{fbound}), showing that, even if the use of ancillas is prohibited, a joint discrimination strategy will still outperform the method of  pairwise elimination.   

The conditions for unambiguous discrimination are sufficient to guarantee perfect deterministic discrimination when the set $\set U$ is a generalized $t$-design.  Indeed, also in the ancilla-free case we show that for generalized $t$-designs there is no difference between probabilistic and deterministic discrimination strategies.     
 Finally, when the gates in $\set U$ form a representation of a group  $\grp G$, one can prove more specific results \cite{thesis}:
\begin{enumerate} 
\item  A perfect discrimination strategy using  $d_A$-dimensional ancilla can be replaced by a perfect ancilla-free discrimination strategy  using
 \begin{eqnarray}\label{deltaN}
\Delta N_{\min}  \le \left\lceil    \frac {  \log d_{A} + \log \sqrt{   |\grp G|}   }{\log  d   }   \right\rceil  
\end{eqnarray}
extra queries.   This result is consistent with the scaling with the number of qubits promised by Eq. (\ref{O}): for a strategy using a large number of ancillary qubits $N_A  \gg   \log\sqrt{ |\grp G|}$, the number of extra queries to the black box scales as $O(N_A)$.   
\item   The query complexity of  ancilla-free discrimination can be upper bounded with an expression involving the maximum entanglement fidelity between pairs of different gates (cf.  Subsection \ref{subsect:groups} for the actual expression).  This is quite surprising because the operational interpretation of the entanglement fidelity is the fidelity between the output states obtained by applying the unitaries on one side of the maximally entangled state, a strategy that is forbidden in ancilla-free gate discrimination.  When the entanglement fidelity is zero, we obtain that the minimum number of queries needed for  ancilla-free discrimination is given by
\begin{equation}\label{toofavourable}
N^{AF}_{\min}   =  \lceil \log_d |\grp G| \rceil \, .
\end{equation}  
In principle, this is the most favourable scaling possible: indeed, with less than  $  \lceil \log_d  |\grp G|\rceil$ queries it would be impossible to  pack $ |\grp G|$ orthogonal vectors in the joint Hilbert space of the systems used  by the discrimination strategy.

\end{enumerate}

Eqs. (\ref{deltaN}) and (\ref{toofavourable}) allow one to quantify the resources needed for protocols of quantum communication  \cite{commun} and decoherence-free encoding \cite{skoti} in the absence of shared reference frames.  Indeed, in these cases $N_{\min}$ is  equal to the number of physical systems needed to construct the ``token state" used in the communication protocol \cite{commun,skoti}. In the case of the alignment protocols \cite{kempe,refframe,bagan,gloves,colorcod,qcmc},  $N^{AF}_{\min}$ is the minimum number of quantum systems that have to be exchanged in order to establish a reference frame.  

\section{Gate discrimination: framework and basic definitions}\label{sec:abstention}

We consider the problem of identifying an unknown unitary gate under the promise that the gate belongs to a finite set $\set U$.  
 For simplicity, we assume that the gates act on a system with Hilbert space $\spc H$ of finite dimension $d< \infty$.  
 Moreover, all throughout the paper we assume that every two unitaries  $U_x ,U_y \in \set U$ are statistically distinguishable, that is, there exists at least one input state, with density matrix $\rho $, such such that 
 \begin{eqnarray}\label{projfaith}   
 U_x  \rho U_x^\dag  \not  =           U_y \rho  U_y^\dag  .  
 \end{eqnarray}    
If this were not the case,  there would be no point in making an experiment to distinguish between $U_x$ and $U_y$, because these two gates would give rise to the same outcome probabilities for every possible experiment, and, therefore, there would be no operational way to tell them apart.


\subsection{Discrimination strategies}
In order to identify the action of the unknown gate, one is allowed to make queries to the corresponding black box and to use them in an arbitrary quantum circuit.  
As long as there is no constraint on the use of ancillas, one can focus  without loss of generality on circuits consisting of pure states and unitary gates, 
 of the form
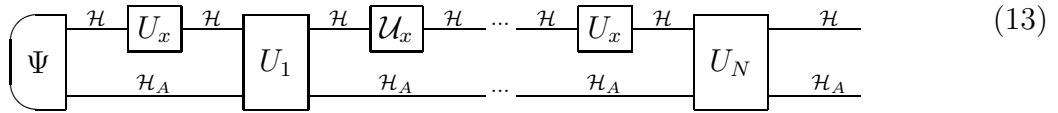
\begin{equation}\label{genstrat}  
\mbox{\Qcircuit @C=1em @R=.7em @! R {
\multiprepareC{1}{   \Psi} &\qw \poloFantasmaCn{\spc H}   &\gate{U_x }  & \qw \poloFantasmaCn{\spc H }  & \multigate{1}{U_1} & \qw\poloFantasmaCn{\spc H}  & \gate{\map U_x}    &  \qw \poloFantasmaCn{\spc H}   & \qw &   \poloFantasmaCn{\dots\quad } &       \qw \poloFantasmaCn{\spc H}  &    \gate{U_{x}}  &\qw \poloFantasmaCn {\spc H}     &\multigate{1}{U_N}   &  \qw &\qw \poloFantasmaCn{\spc H}&\qw\\
\pureghost{\Psi} &\qw &\qw  \poloFantasmaCn{\spc H_{A}}  & \qw & \ghost{U_{1}}  &  \qw  & \qw   \poloFantasmaCn{\spc H_{A}} & \qw &\qw  &\poloFantasmaCn{ \dots\quad }  & \qw   &\qw   \poloFantasmaCn{\spc H_{A}}& \qw  &\ghost{U_N}  &\qw & \qw \poloFantasmaCn{\spc H_{A} } &\qw \\  } }    \qquad   \qquad 
 \end{equation}  
 \medskip
where
\begin{enumerate}
\item  $|\Psi\>  \in\spc H \otimes \spc H_A$ is the joint state state of the input of $U_x$ and of an ancilla with Hilbert space $\spc H_A$
\item $U_t$ is a unitary gate representing a joint evolution of the system and the ancilla   at the time step $t\in\{1,\dots, N\}$ (the unitary $U_N$ is added just for convenience of notation). 
\end{enumerate}

Once the input state $|\Psi\>$ and the unitaries $(U_t)_{t=1}^N$ have been chosen, identifying the unitary $U_x$ is equivalent to identifying the output state
\begin{eqnarray}\label{outstate}
|\Psi_x\>  :=     \left[\prod_{n=1}^N     U_n(  U\otimes I_{A} ) \right]~ |\Psi\> \, .
\end{eqnarray} 
To this purpose, one has to perform a suitable quantum measurement, described by a joint POVM $(P_y)_{y\in\set Y}$.  Here we allow for measurements with a set of outcomes   $\set Y  =   \set X  \cup  \{  ?\}$, including an inconclusive outcome $y=?$, which corresponds  to the case when the experimenter abstains from producing a guess \cite{bergouhill1}. 
Among all possible strategies, the deterministic ones are those for which   
the inconclusive outcome never occurs, namely $P_?  =  0$.  
\medskip

\subsection{Optimal, error-free, unambiguous, and perfect discrimination}

As a figure of merit for gate discrimination, we choose the probability that that the unknown gate is identified correctly, provided that the measurement does not output the inconclusive outcome $y  = ?$.   This  probability is given by  
\begin{eqnarray}\label{pN}
p_{N}   : =   \frac{ \sum_{x\in\set X}   p_N(x|x)  ~ p_x }{\sum_{  x,y\in \set X }    p_N(y|x) ~ p_x}.       
\end{eqnarray}
where  $p_x$ is the prior probability of $U_x$ and $p_N (y| x)  = \< \Psi_x|  P_y |\Psi_x\>$  is the conditional probability of the measurement outcome $y$ given that the gate is  $U_x$ and that $N$ queries are used.  

The optimal discrimination strategy is the one that maximizes the success probability $p_{N}$. We denote the corresponding probability by $p_N^{\max}$.  Note that, in general, a deterministic strategy may not be able to reach the value $p_N^{\max}$: in order to achieve the optimal performances one may be forced to have  an inconclusive outcome. For this reason, the probability $p_N^{\max}$ is an upper bound on the maximum probability of success over deterministic strategies, which is the quantity normally considered in minimum-error discrimination.

In this paper we will be particularly interested in discrimination strategies that are \emph{error-free}, in the sense that they never misidentify the gate ($p_N=1$).   Note that the error-free condition $p_N=1$ is much weaker than it may seem at first sight:  this can be seen in the example of  the qubit gates 
\begin{equation*}
U_0  =  (I+ Z)/\sqrt 2  \qquad  U_k  =  \cos (2\pi/K)  I  +  i\sin(2\pi/K)  X \quad k=  1,\dots , K,
\end{equation*}
where  $K$ is an arbitrary integer number.  In this case, one can achieve success probability $p_N  =1$ by applying the unknown unitary on one side of the maximally entangled state $|\Phi^+\>  = ( |0\>|0\>  +  |1\>  |1\>)/\sqrt 2 $ and by measuring the output state with the POVM given by
\begin{equation*}
P_0  =  (Z \otimes I)  |\Phi^+\>\<  \Phi^+|   (Z^\dag \otimes I)  \qquad  P_k  =0 \quad \forall k = 1,\dots, K  \qquad  P_?  =  I  - P_0.    
\end{equation*}
Clearly, \emph{when the inconclusive result does not occur}, the unknown gate has been identified with certainty: indeed,      the outcome $0$ can only occur when the gate is $U_0$. 

In some situations, having a  discrimination strategy that detects only one gate and aborts otherwise may not be useful.  Instead, one may require that every gate in the set $\set U$ have a non-zero probability of being identified.    We say that a discrimination strategy achieves \emph{unambiguous discrimination} if it is error-free ($p_N = 1$) and, in addition, $p(x|x)  >  0$ for every $x$.   For example, the two unitaries $U_0  =I$ and $U_1  =\exp[i  \theta Z] $ can be distinguished unambiguously by preparing the input state $|+\>  =(|0\>  +  |1\>)/\sqrt 2$ and by measuring the POVM defined by 
\begin{equation*}
P_0  =  \frac{  U_1  |-\>\<  -  |  U_1^\dag}{1+\cos (\theta/2)} \qquad P_1  =  \frac{  U_0  |-\>\<  -  |  U_0^\dag}{1+\cos (\theta/2)}  \qquad P_?  =  I -  P_0  -  P_1, 
\end{equation*}
with $  |-\>  =  (  |0\>  -  |1\> )/\sqrt 2$.   This strategy is error-free $ p_N  = 1$ and both gates have the chance of being detected: in this particular case, one has $p( 0|0)  =  p(1|1)  = [\sin(\theta/2)]^2/[1+\cos(\theta/2)] $. 

Note, that the definition of unambiguous discrimination does not include any requirement  on the probability of the inconclusive outcome, which in principle can be arbitrarily close to 1.  In some situations, this feature can undermine the usefulness of the discrimination scheme.  On the opposite end, one can restrict the attention to discrimination strategies such that the probability of the inconclusive outcome is equal to 0.  We refer to these strategies as \emph{perfect discrimination strategies}.   

\subsection{Basic facts about error-free and unambiguous discrimination}
 Error-free and unambiguous discrimination can be nicely characterized in terms of linear independence: 
\begin{Theo}\label{theo:unambiguous}
The unitaries in $\set U $ can be discriminated in $N$ queries 
\begin{enumerate}
\item in an error-free way  if and only if there exists a unitary $U_{x_0}$ that is not a linear combination of the other unitaries  in  $\set U$ 
\item in an unambiguous way if and only if the unitaries $(U^{\otimes N}_{x})_{x\in\set X}$ are linearly independent.
\end{enumerate}
\end{Theo}

\noindent  The equivalence between unambiguous gate discrimination and linear independence of the unitaries was previously observed in Ref. \cite{chefles2} in the case of \emph{parallel strategies}, i.e. strategies where the $N$ queries are applied in parallel to a suitable multipartite state $ |\Psi\> \in\spc H^{\otimes N} \otimes \spc H_A$, thus producing the output state   $|\Psi_x\>  =  \left(U_x^{\otimes N}  \otimes I_A\right)|\Psi\>$, as in figure
 \begin{equation}\label{parastrat} 
\qquad \qquad \qquad\qquad \mbox
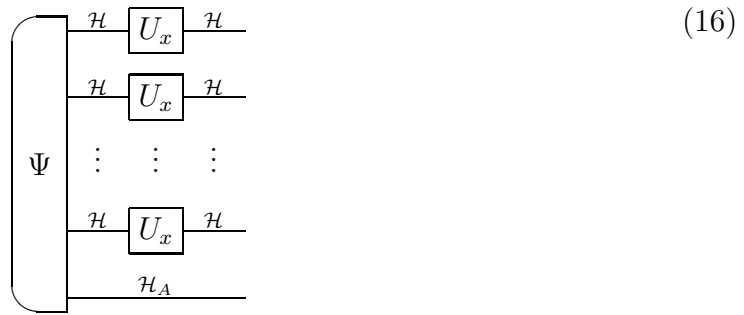
{\Qcircuit @C=1em @R=.7em @! R {
\multiprepareC{4}{ \Psi} &\qw \poloFantasmaCn{\spc H}   &\gate{U_x }  & \qw \poloFantasmaCn{\spc H }  &\qw  \\   
\pureghost{   \Psi} &\qw \poloFantasmaCn{\spc H}   &\gate{U_x }  & \qw \poloFantasmaCn{\spc H }  & \qw\\
\pureghost{   \Psi} &  \vdots &\vdots  &  \vdots &  \\
\pureghost{   \Psi} &\qw \poloFantasmaCn{\spc H}   &\gate{U_x }  & \qw \poloFantasmaCn{\spc H }  &\qw \\
\pureghost{   \Psi} &\qw   &\qw   \poloFantasmaCn{\spc H_A}& \qw  & \qw  
} }   
 \end{equation}  
Parallel strategies are a special case of the strategies of Eq. (\ref{genstrat}), where one has the freedom to apply the $N$ queries at different time steps in a quantum circuit.  The parallel strategies of Eq. (\ref{parastrat}) can be recovered as a special case from the general strategies of Eq. (\ref{genstrat})  by setting the gates $(U_t)_{t=1}^N$ to be suitable swap gates, as in the following example 
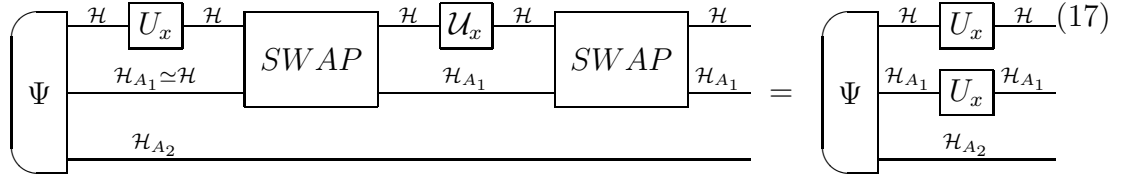
\begin{equation}\label{parastrat} 
\mbox{\Qcircuit @C=1em @R=.7em @! R {
\multiprepareC{2}{\Psi} &\qw \poloFantasmaCn{\spc H}   &\gate{U_x }  & \qw \poloFantasmaCn{\spc H }  & \multigate{1}{SWAP} & \qw\poloFantasmaCn{\spc H}  & \gate{\map U_x}    &  \qw \poloFantasmaCn{\spc H}   & \multigate{1}{SWAP}  &  \qw \poloFantasmaCn{\spc H}   &\qw   &  \\
\pureghost{\Psi} &\qw &\qw  \poloFantasmaCn{\spc H_{A_1}  \simeq \spc H}  & \qw & \ghost{SWAP}  &  \qw  & \qw   \poloFantasmaCn{\spc H_{A_1}} & \qw&\ghost{SWAP}  & \qw \poloFantasmaCn{\spc H_{A_1}}  &\qw     &  = \\ 
\pureghost{\Psi} &\qw &\qw  \poloFantasmaCn{\spc H_{A_2}}  & \qw & \qw  &  \qw  & \qw    & \qw &\qw &\qw&\qw  &
}}  
\mbox{    \Qcircuit @C=1em @R=.7em @! R {
&\multiprepareC{2}{ \Psi} &\qw \poloFantasmaCn{\spc H}   &\gate{U_x }  & \qw \poloFantasmaCn{\spc H }  &\qw  \\   
&\pureghost{   \Psi} &\qw \poloFantasmaCn{\spc H_{A_1}}   &\gate{U_x }  & \qw \poloFantasmaCn{\spc H_{A_1} }  & \qw\\
&\pureghost{   \Psi} &\qw   &\qw   \poloFantasmaCn{\spc H_{A_2}}& \qw  & \qw  
} }   
 \end{equation}  
Theorem \ref{theo:unambiguous} has an important consequence:  it implies that error-free discrimination and unambiguous discrimination discrimination can be parallelized: 
\begin{Corollary}[Parallelization of error-free and unambiguous discrimination]\label{cor:aaa}
If the gates $\set U $ can be  distinguished unambiguously (respectively, in a error-free fashion) with $N $ queries, then they can be distinguished unambiguously (respectively, in a error-free fashion) using the $N$ queries in parallel.
 \end{Corollary}

In other words, the identification of the gate can be achieved in the shortest possible number of time-steps: in the problem of error-free and unambiguous discrimination the time resource can be completely replaced by spatial resources. 
This fact is also useful as a technical tool: it implies that the query complexity of error-free/unambiguous discrimination---defined as  minimum number $N_{min}$ needed to unambiguously identify a gate in $\set U $---does not change if one restricts to  parallel strategies.  Note, however, that general sequential strategies can help in reducing the probability of the inconclusive result.   This fact is well illustrated by the  example of Grover's algorithm: 

\begin{Example}[Discrimination of Grover's oracles]
Grover's algorithm is designed to identify a unitary gate in the set $\set U$  containing the gates  $U_x  =   I-  2 |x\>\<  x|$, $x\in \set X  =  \{1,\dots, d\}$.  Clearly, the gates $\set U$ are linearly independent for every $d>2$ and therefore they can be unambiguously discriminated in a single query, as originally observed  by Chefles {\em et al} in Ref. \cite{chefles2}.  However, the probability of unambiguous discrimination in a single query must be necessarily low.   One way to see it is the following: as showed by Brassard, Hoyer, Boyer, and Tapp \cite{brassard} and Zalka  \cite{zalka}, Grover's algorithm cannot be efficiently parallelized:  there is no deterministic parallel strategy that can achieve in $N$ queries the same probability of correct gate identification as in Grover's algorithm.  Now, if the probability of unambiguous discrimination were sufficiently large,  one could run different rounds of unambiguous discrimination, and use this fact to construct an efficient parallel strategy.    In the case of search in a large database ($d\gg 1$), denoting by $N_G$  is the number of queries needed by Grover's algorithm to achieve maximum  probability of correct gate identification,  one can show that the probability of unambiguous discrimination in one query must be upper bounded by $O(\log  N_G/N_G)$.    
\end{Example}


\section{General bounds on the query complexity of unambiguous gate discrimination}\label{sec:complexity}
The possibility of parallelizing unambiguous gate discrimination, established by theorem \ref{theo:unambiguous}, leads immediately to general bounds on the query complexity. These bounds do not assume any structure of the set of unitaries $\set U$ and can typically be improved when more information about $\set U$ is available.  

\subsection{Lower bound}

Here we give a lower bound to the number of queries that are necessary to for unambiguous gate identification.
\begin{Proposition}[Dimensional bound]\label{bnd:lower} 
If the the gates  in   $\set U  $ can be unambiguously discriminated using $N$ queries, then
\begin{equation}\label{sud}
|{\mathsf U}| \le  
\left(   \begin{array}{c}
  N + d^2- 1 \\
 d^2-1 
 \end{array}
 \right)  .
\end{equation}
\end{Proposition}

If we do not impose any structure on the set of unitaries $\set U $, then the bound of Eq. (\ref{sud}) is the best we can hope for.  Indeed, for any fixed Hilbert space dimension $d$ and for every size $| \set U|$ we can always find a set of unitaries $\set U $  such that the minimum number of queries needed to unambiguously identify a gate in $\mathsf U$ is exactly  the minimum $N$ compatible with Eq. (\ref{sud}) (cf. the proof in the Appendix).
  
Eq. (\ref{sud}) can be used to provide an easy lower bound on the query complexity:   Combining it with  the inequality 
\begin{equation}\label{binombound}
\left(   \begin{array}{c}
 N +  k \\
 k 
 \end{array}
 \right) <   (N+1)^{k}, 
\end{equation}
we obtain the bound   $N  >   |\set U|^{\frac 1 {d^2 -1}}  -1 $, which is a necessary condition for unambiguous discrimination with $N$ queries. The bound is not tight, but  provides the right scaling with $|\set U|$ in the regime when $d$ is fixed and $N$ is large compared to $d^2$.   Indeed, in
this case one has
\begin{equation*}
\left(   \begin{array}{c}
 N +  d^2-1 \\
   d^2-1 
 \end{array}
 \right) =   \frac{N^{d^2-1}}{(d^2 -1)!}  +   O (  d^2/N),  
\end{equation*}
which means that the scaling of the tight bound associated to  Eq. (\ref{sud}) is actually $  N =   \Omega\left(   |\set U|^{\frac 1 {d^2-1}}\right)$. 

\subsection{Upper bounds} 

An upper bound on the query complexity can be obtained by observing that the number of linearly independent unitaries in $\set U_N $  grows at least linearly with $N$, a fact that can be proved using an earlier result by Chefles \cite{Chefles}:  
\begin{Proposition}[Linear bound]\label{bnd:linear}  The query complexity of unambiguous discrimination of the gates in  $\set U$ is upper bounded by  
\begin{equation}\label{linear}
N_{\min}  \le   |\mathsf U|  +1 - \dim (\mathsf U) .     
\end{equation}  
\end{Proposition}

\medskip 
In general, the bound of Eq. (\ref{linear}) can be achieved: for every fixed Hilbert space dimension $d$ and for every fixed cardinality $|\mathsf U|$ we can find a set of unitaries such that $N_{\min}  =   |\mathsf U|  - \dim (\mathsf U)  + 1$.    This can be seen  in the following

\begin{Example}[Unambiguous discrimination of discrete phase-shifts]  Consider the problem of identifying an unknown phase shift
\begin{eqnarray*}U_x  :  =  \omega^x ~ |1\>\<  1|   +   (I  -|1\>\<  1|   ) \qquad   \omega:  =  e^{\frac {2\pi i}{|\mathsf X|}},
\end{eqnarray*} 
with $x= 1,\dots,  |\mathsf X|$. In this case the number of linearly independent unitaries in $(U_x^{\otimes N})_{x\in\set X}$ is exactly equal to $   N +1$, as it can be seen from the fact that the unitaries   $(U_x^{\otimes N})_{x\in\set X}$ are in one-to-one correspondence with the vectors of their eigenvalues, given by $  (  v_x)_{x\in \set X} \subset  \Cmplx^{N+1}$ where $v_x  :=  ( 1, \omega, \omega^2, \dots, \omega^N )^T$.  Since the number of linearly independent unitaries in $(U_x^{\otimes N})_{x\in\set X}$ is $N+1$, the minimum number needed for unambiguous discrimination is exactly $N_{\min} =   |\set X|-1  =  |\set U|  -  \dim (\set U)  + 1 $.  
  \end{Example}
  
 \medskip 
Another example where the bound of Eq. (\ref{linear}) gives the exact value is the identification of a ``shift-and-multiply" gate: 
\begin{Example}[Unambiguous discrimination of shift-and-multiply gates]\label{ex:shiftmult} Consider the problem of identifying a shift-and-multiply gate
\begin{equation}\labell{SchiftAndMultiplyRep}
U_{pq}=S^p M^q 
\qquad (p,q) \in \mathbb Z_d \times \mathbb Z_d ~,
\end{equation} 
where $S = \sum_{k=1}^{d} |(k + 1)  {\rm mod}~ d \>\<k|$ and $M=
\sum_{k=1}^{d} e^{(2\pi i k)/d} |k\>\<k|$. 
In this case, the unitaries $(U_{pq})_{(p,q) \in  \mathbb Z_d \times \mathbb Z_d}$ are linearly independent, and therefore  the bound gives $N_{\min} =  1$. Note that,  in fact, the unitaries are orthogonal in the Hilbert-Schmidt product, and, therefore, an unknown unitary $U_{pq}$ can be identified perfectly and deterministically, as in the dense coding protocol \cite{dense}. 
\end{Example}

\medskip

Proposition \ref{bnd:linear}  provides an estimate of $N_{\min}$ that is always better than the number of pairwise tests $ |\mathsf U|  -1$  that would be needed to identify a gate in $\set U$ with the method of pairwise eliminations outlined in \cite{Acin,EntImproves}.     Note however that Eq. (\ref{linear}) only ensures \emph{unambiguous} discrimination, while the pairwise elimination method ensures \emph{perfect} discrimination.    In the next Section we will see that the distinction between unambiguous and perfect discrimination disappears when the gates in $\set U$ form a group representation, or, more generally, a generalized $t$-design.  

Before adding more structure on the set $\set U$, we give here a second upper bound that often yields a better estimate than Proposition \ref{bnd:linear}.   
To state the bound we introduce the \emph{minimax fidelity} of the unitaries in $\set U$, defined as  
\begin{eqnarray*}
F_{\set U}  &:  =\min_{  |\Psi\> \in  \spc H  \otimes \spc H , |\!|  \Psi |\!|  = 1} \quad\max_{x, y \in \set X,  x\not =y}     \left |  \<\Psi|    (  U_x^\dag   U_{y}   \otimes I)  |\Psi\> \right|  .  \end{eqnarray*}
The minimax fidelity quantifies the pairwise distinguishability of the gates in $\set U$ when single-shot ancilla-assisted strategies are used.   Clearly, if  $F_{\set U}  = 0$, the unitaries can be perfectly distinguished in one shot using a suitable input state.   Note also that, under the standing assumption of this paper,  $F_{\set U}$ must be strictly smaller than $1$: indeed, the distinguishability condition of Eq. (\ref{projfaith}) implies that for every two distinct unitaries $U_x$ and $U_y$ there exists at least an input state $|\psi\>$ such that $\left|\<  \psi  |  U_x^\dag  U_y|\psi\>\right|^2<1$.      
In terms of the minimax fidelity, we have the following
\begin{Proposition}[Fidelity bound]\label{theo:fidelity}
The query complexity of unambiguous discrimination of the gates in  $\set U$ is upper bounded as  
\begin{equation}\label{fide}
N_{\min}  \le    \left \lfloor \frac{\log   (|\set U| -1)}{\log\left( F^{-\frac 12}_{\set U}\right) }  \right \rfloor +1 .      
\end{equation}  
\end{Proposition}

The proof of the bound is based on a simple observation about unambiguous state discrimination, which is interesting \emph{per se}: 
\begin{Lemma}\label{lemma:statedisc}  
Let $(  |\psi_x\>)_{x\in \set X}  \in \spc H$ be a set of pure states and let  $F : = \max_{ x\not = y}   |  \<  \psi_x|\psi_y\>|^2$ be the maximum fidelity between two distinct states in the set.     
If $ F^{N/2}   <   1/(|\set X|-1)$, then the states $(  |\psi_x\>^{\otimes N})_{x\in\set X}$ are linearly independent, and, therefore, unambiguously distinguishable.     
\end{Lemma}

For qubits, this simple observation gives exactly the  minimum number  of copies needed for unambiguous discrimination of the states in a \emph{symmetric informationally-complete (SIC) POVM} \cite{sic1,sic2}. Recall that a SIC-POVM in dimension $d$  is a set of $ d^2$ unit vectors with the property that the overlap between any two distinct vectors is the same:
\begin{equation*}
|\<  \psi_x| \psi_y \> |^2  =  \frac {1}{d+1}      \qquad\forall x\not =  y. 
\end{equation*}  
\medskip 
In general, for $d^2$ pure states one can easily see from dimensional arguments that unambiguous discrimination requires at least  $3$ copies  (cf. the lower bound in Ref. \cite{Chefles}).  
 On the other hand, Lemma \ref{lemma:statedisc} shows that for qubits  $N=3$ copies are sufficient, thus implying that $N=3$ is actually the minimum number of copies needed for unambiguous discrimination of a SIC-POVM.     For general $d$-dimensional systems,  Lemma \ref{lemma:statedisc} gives  the upper bound  $N\le4 $, almost matching the dimensional lower bound \cite{comment}, and thus showing that the number of copies does not scale up with the dimension of the system.  
  The exact value of the minimum number of copies is equal to $N_{\min}  = 3$ for all the known examples of SIC-POVMs, except for the one example of SIC POVM in dimension $d=3$, which actually requires $N_{\min}  =  4$  copies \cite{comment}. 
  
\medskip 

Let us now comment on the tightness of the fidelity bound for  gate discrimination.   The bound gives good estimates when $F_{\set U}$ is close to zero (in particular, in the extreme case where $F_{\set U}=0$ it predicts correctly $N_{\min}  =1$).     However, it tends to overestimate $N_{\min}$  when $F_{\set U}$ approaches $1$.  To understand this fact, note that for $F  \ge  1-  \epsilon$, the estimate of $N_{\min}$ becomes 
\begin{equation*}
N_{\min}  \le  \frac{ 2 \ln(  |\set U|  -1)}{\epsilon}~,
\end{equation*}  
having chosen the logarithm in base $e$.
Now, two unitaries can have fidelity arbitrarily close to 1 and still be linearly independent. For example, the unitaries $U_0  =  I$ and $U_1  = e^{i\theta Z}$ are linearly independent, and their fidelity is $F=  \cos \theta$.  This implies that for every set $\set U$ containing these two unitaries one has $F_{\set U}  \ge  \cos \theta $ and, therefore,  the fidelity bound gives $N_{\min}     \le  \frac{ 4 \ln(  |\set U|  -1)}{\theta^2}$, where 
the r.h.s. can be arbitrarily large when  the angle $\theta$ is small. Hence, for $F_{\set U} \approx 1$ the fidelity bound can be arbitrarily far from the correct value (think of the case when the unitaries $\set U$ are linearly independent, and the correct value is  $N_{\min}  = 1$).   Another example of the gap between the value of the upper bound and true value of $N_{\min}$  for $F_{\set U}  \approx 1$ is illustrated in the following

\begin{Example}[Permutation gates]\label{ex:permu} 
Consider 
identification of an unknown permutation gate
\begin{eqnarray}\label{permu}
U_{\pi}  =\sum_{k=1}^d    |\pi(k)\>\<  k|  ,    
\end{eqnarray}
where $\pi $ is an element of the  permutation group $S_d$. 
In this case it is clear that the unitary $U_\pi$ can be perfectly identified with  $d$ queries (applying $U_\pi$ to all the $d$ vectors in the computational basis we can surely identify the permutation $\pi\in S_d$).  
One the other hand, applying the unitary $U_\pi$ on a maximally entangled state gives the bound $F_{\set U}  \ge   \left(\frac { d-2}  d\right)^2$, which inserted in  the fidelity bound gives  $N_{\min}  \le   \log (d!)/\log[d/(d-2)]  =  O   (d^2  \log d)$, which is off by a factor $d \log d$ from the actual value. 

 \end{Example}
 

\section{Discrimination of generalized $t$-designs}\label{sec:designs} 

Here we impose additional structure on the set  of  gates  $\set U $. 
Our analysis  includes the case where  the set $\set X$ labelling the gates in $\set U$ is a finite group and $x\mapsto U_x$ is a  representation of $\set X$.   
Also, it includes the case where the unitaries $\set U $ from a unitary $t$-design \cite{tdesign1,tdesign2,tdesign3,tdesign4}.   
In order to treat these two cases in a unified way, we introduce the notion of \emph{generalized $t$-designs}.   
 For  the discrimination of generalized $t$-designs we will show the following properties  
\begin{enumerate}
\item among all possible discrimination strategies using $N\le t $ queries, the deterministic strategies using all queries in parallel maximize the probability of correct gate identification 
\item for strategies using $N\le t$ queries, there is no difference between error-free, unambiguous, and perfect discrimination. 
\item   the maximum probability of correct gate identification with $N \le t$ queries has a simple analytic formula. 
\end{enumerate}

\subsection{Generalized $t$-designs: definition and characterization }
The notion of unitary $t$-design plays an important role in quantum information, with applications essentially in all protocols that require to extract a random gate from the uniform distribution over all possible unitaries \cite{tdesign1,tdesign2,tdesign3,tdesign4}.   Unitary $t$-designs are defined as follows:  Consider an ensemble $ ( U_x,p_x)_{x\in\set X}$ where $U_x$ is a unitary and $p_x$ is a probability.   
The ensemble is a unitary $t$-design iff
\begin{equation*}\sum_{x\in \set X}   p_x    U_x^{\otimes t}  \otimes \overline U_x^{\otimes t}  =  \int  {\rm d}U    ~  U^{\otimes t}  \otimes \overline U^{\otimes t},
\end{equation*}   
where $\overline U$ denotes the complex conjugate of the matrix $U$ and the integral in the l.h.s. runs over the normalized Haar measure on the unitary group $U(d)$.

We will now generalize the above definition considering, instead of the full unitary group, a smaller group of unitary gates:  
\begin{Def}[Generalized $t$-designs] Let $U:  g\mapsto U_g$ be a representation of a group $\grp G$ and let $\set X$ be a subset of $\grp G$.   An ensemble $(U_x, p_x)_{x\in\set X}$  is a \emph{generalized $t$-design} iff 
\begin{eqnarray}
 \sum_{x\in\set X}   p_x    ~  U_x^{\otimes t}   \otimes  \overline U_x^{\otimes t}  =  \int  {\rm d}g ~     U_g^{\otimes t}   \otimes  \overline U_g^{\otimes t} ,
\end{eqnarray} 
where $\int {\rm d}g f(g)$ denotes the integral of $f$ with respect to  the normalized Haar measure.  
\end{Def}

Note that, by definition, every  generalized $t$-design is also a generalized  $(t-1)$-design.  Of course, a special case of generalized $t$-design is obtained by taking $\grp G$ to be a finite group and $\set X  =  \grp G$: 
\begin{Example}   The ensemble consisting of unitary gates in a representation of a finite group $\grp G$, randomly sampled with  uniform probability distribution $p_g = 1/ |\grp G|$, is a generalized $t$-design for every $t$.  
\end{Example}

The definition of $t$-designs is extremely convenient, because it allows to easily transfer properties of groups to finite sets of quantum gates. In the next sections we will use this trick to prove strong properties of gate discrimination in the case of generalized $t$-designs.

\subsection{Optimal discrimination of generalized $t$-designs}  

We start from general result about optimal probabilistic gate discrimination.  Precisely, we 
 show  that the maximum probability $p^{\max}_N$ of correct  gate identification  can be always achieved with a parallel strategy. In addition, we give an analytic expression for $p_N^{\max}$.
 
\begin{Theo}[Optimal probabilistic gate discrimination]\label{theo:optprob}  
For every choice of prior probabilities,  the maximum success probability $p_N^{\max}$ [cf.Eq. (\ref{pN})] is achieved by applying the $N$ queries in parallel on an entangled state.   
The probability $p_N^{\max}$ is given by
\begin{equation}\label{pmax}
p^{\max}_{ N } =   \max_{x\in\set X}  ~p_x   \<\!\<  U_x  |^{\otimes N} ~  R_N^{-1} ~  |U_x\>\!\>^{\otimes N} ,             
\end{equation}
with  $|U_x\>\!\> :=  (U_x\otimes I)|I\>\!\>$, $|I\>\!\>  :=  \sum_{n=1}^d  |n\> |n\>$, 
$R_N :=   \sum_{x\in\set X}  p_x ~  \left ( |U_x\>\!\>\<\!\<  U_x  | \right)^{\otimes N}$, and $R_N^{-1}$ being the inverse of $R_N$ on its support.
\end{Theo}

\medskip
The explicit formula of Eq. (\ref{pmax}) is useful even if one is interested in deterministic strategies, rather than probabilistic ones.  Indeed, by definition $p_N^{\max}$  provides an upper bound to the probability of success of arbitrary deterministic strategies.   In some cases, one may even be able to achieve the upper bound with a deterministic strategy.  This is actually the case for generalized $t$-designs:  
\begin{Theo}[Optimal gate discrimination for generalized $t$-designs]\label{theo:design}  Let $(U_x,p_x)_{x\in\set X}$ be a generalized $t$-design.  
Then, the maximum of the probability of correct discrimination  over all  probabilistic strategies consisting of $N\le t$ queries  is   
\begin{eqnarray}\label{pmaxdesign}
p^{\max}_{N}  =    \dim  \left( \set U_{N} \right)  ~  \max_{x \in\set X}  p_x  
\end{eqnarray}
For uniform prior $p_x  = 1/|\set U  |$,   the maximum probability $p_N^{\max}  =   \dim(\set U_{N})/|\set U|$ can be achieved by a deterministic strategy that uses the $N$ queries in parallel. \end{Theo}  


 



\medskip 

The general result of theorem \ref{theo:design} is well illustrated by the case of discrete phase shifts:
\begin{Example}[Discrete phase shifts]\label{ex:discphase}
Consider problem of identifying a  discrete phase-shift  gate
\begin{eqnarray}\label{discphase} 
U_x   =   \sum_{y=0}^{d-1}   \omega^{ x y } ~  |y\>\<  y| \qquad \omega  =  e^{\frac{2\pi i}{|\set X|}} , \end{eqnarray}
 $U_x$  chosen at random with uniform probability $p_x  =  1/|\set U|$.    
  Here $U:  x \mapsto U_x$ is a unitary representation of the Abelian group $\set X  = \mathbb Z^{K}$, and, therefore, it is  a generalized $t$-design for every $t$.  
Now, the number of linearly independent unitaries in $\set U$ is $d$. Hence, the probability of correct identification of a unitary with a single query is $p_1^{\max}  =  d/|\set U|$.    
Similarly, the number of linearly independent unitaries in $\set U_N$ is $  \min\{ N (d-1) +1,   |\set U|\}$, and therefore, Eq. (\ref{pmaxdesign}) gives  
\begin{eqnarray} 
p_N^{\max}  =  \frac  { Nd  - N  + 1 }{|\set U|} \qquad N  \le \frac{|\set U|-1}{d-1}.  
\end{eqnarray}
\end{Example}

\subsection{Perfect discrimination of generalized $t$-designs}  
From now on, we restrict our attention to generalized $t$-designs with \emph{uniform} probability distribution $p_x  =  1/|\set X|$.  Since there is no ambiguity, we will just refer to them as ``generalized $t$-designs".    An immediate consequence of Theorem \ref{theo:design} is that for generalized $t$-designs there is no difference between error-free, unambiguous, and perfect gate discrimination: 
\begin{Corollary}\label{theo:equivalence}  
If the unitaries  $(U_x)_{x\in \set X }$  form a generalized $t$-design, then the following are equivalent:  
\begin{enumerate} 
\item error-free discrimination is possible with  $N\le t$ queries 
\item  unambiguous discrimination is possible with $N\le t$ queries 
\item perfect discrimination is possible in $N\le t$ queries. 
\end{enumerate}
In particular, for a generalized $|\set U|$-design there is no difference between error-free, unambiguous, and perfect discrimination.
\end{Corollary}     

For generalized $t$-designs the evaluation of the query complexity of perfect discrimination is reduced to the simpler problem of evaluating the query complexity of unambiguous discrimination.  In particular, the bounds in Propositions \ref{bnd:lower}, \ref{bnd:linear}, and \ref{theo:fidelity} become automatically bounds on the query complexity of perfect discrimination.  

\section{Bounding the dimension of the ancilla}\label{sec:ancilla}
In addition to the query complexity, it is useful to bound  the size of the ancilla needed for gate discrimination \cite{entest,chen}.  Here we show that the size of the optimal ancilla scales at most polynomially with the number of queries $N$.   We prove this result as a particular case of a more powerful statement about discrimination of unitaries picked from a group representation. 

\begin{Proposition}\label{bnd:ancilla}  
Let $ U  : g\mapsto U_g$ be a representation of a group $\grp G$ and let $\set U$ be a subset of $(U_g)_{g\in\grp G}$   Then, the minimum dimension of the ancilla needed for unambiguous discrimination of the gates $\set U$  in $N$ queries is upper bounded by 
\begin{eqnarray}\label{dimanc}
d_{A,N}^{\min}  \le   \max_{\mu  \in  {\rm Irr}  \left(U^{\otimes N}\right)} \left\lceil \frac{d_\mu}{m_\mu} \right\rceil ,
\end{eqnarray}
where the maximum runs over the set ${\rm Irr}  \left(U^{\otimes N} \right)$of irreducible representations contained in the decomposition of $U^{\otimes N}$, $d_\mu$ and $m_\mu$ are  the dimension and the multiplicity of the irreducible representation $\mu$ (see the Appendix for some background information on representation theory).
\end{Proposition}

\medskip

Proposition \ref{bnd:ancilla} has many useful consequences.  The first is that ancillas are not needed for the unambiguous discrimination of commuting unitaries, a fact that was noted in Ref. \cite{chefles2}  for parallel discrimination strategies.

 \begin{Corollary}  If the unitaries in  $\set U$ commute, then  no ancilla is needed for unambiguous discrimination.  
 \end{Corollary}
 The proof is immediate:  a commuting set of unitaries is a subset of the Abelian group of all unitaries diagonal in a given basis. Since in this case the irreducible subspaces are one-dimensional, Eq. (\ref{dimanc}) gives $d^{\min}_{A,N}  \le 1$, which means that no ancilla is required. 

\medskip 

As anticipated, another consequence of Proposition \ref{bnd:ancilla} is the fact that, no matter which set of gates $\set U$ we are considering, the dimension of the ancilla needed for discrimination in $N$ queries cannot grow faster than a polynomial in $N$. 

\begin{Corollary}  
The minimal dimension of the ancilla needed for unambiguous discrimination  with $N$ queries is upper bounded by 
$d_{A,N}^{\min}  \le \left(     \begin{array}{c}  N  +  d  -1  \\  d-1  \end{array}  \right)$. 
\end{Corollary}

The proof is provided in the Appendix. Bounding the binomial with  Eq. (\ref{binombound}), we have that the dimension of the ancilla is upper bounded by $(N+1)^{d-1}$.   In other words, the number of ancillary qubits needed for unambiguous discrimination scales at most as the logarithm of the number of queries.    

Another application of Proposition \ref{bnd:ancilla} can be found when the group $\grp G$ is finite.  In this case, one can provide an upper bound on the size of the ancilla that is independent of the number of queries:  
\begin{Corollary}\label{cor:sqrtg}   The minimum dimension of the ancilla needed for unambiguous discrimination of the gates $\set U \subseteq  (U_g)_{g\in \grp G}$  is upper bounded by  $d^{\min}_{A,N}  \le \sqrt {|\grp G|}$, independently of the number of queries. 
\end{Corollary}
The bound follows from the fact that, for a finite group $\grp G$, the dimensions of the irreducible subspaces are upper bounded by $\sqrt{\grp G}$ (cf. the background on representation theory provided in the Appendix).  Corollary \ref{cor:sqrtg}  is useful to give a quick estimate of the size of the ancilla. Such an estimate is actually tight in the example of the shift-and-multiply gates $U_{pq}$ (cf. Example  \ref{ex:shiftmult}), which form a representation of the group $\grp G  =\mathbb Z_d\times \mathbb Z_d$.  Since  the size of the group is $|\grp G|  =  d^2$, the bound gives $d^{\min}_{A,N}   \le    d$.  In other words, this means that unambiguous discrimination can be achieved with an ancilla that is of the same size of the input system.    This is exactly what is done by the dense coding protocol  \cite{dense}.

\section{Ancilla-free gate discrimination}\label{sec:noancilla}  

We conclude the paper by briefly discussing  parallel discrimination strategies that do not use ancillas. These strategies involve the preparation of a multipartite input state $|\Psi\>  \in\spc H^{\otimes N}$ and on the application of the unknown gate on each of the $N$ systems, thus obtaining the output state $  |\Psi_x\>  =  (U_x^{\otimes N})|\Psi\>$, as in figure \begin{equation}\label{ancillafree} 
\qquad \qquad \qquad\qquad \mbox
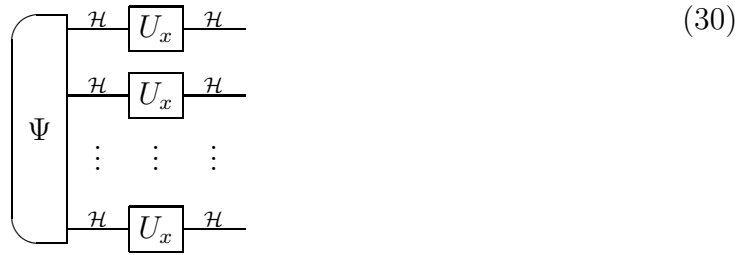
{\Qcircuit @C=1em @R=.7em @! R {
\multiprepareC{3}{ \Psi} &\qw \poloFantasmaCn{\spc H}   &\gate{U_x }  & \qw \poloFantasmaCn{\spc H }  &\qw  \\   
\pureghost{   \Psi} &\qw \poloFantasmaCn{\spc H}   &\gate{U_x }  & \qw \poloFantasmaCn{\spc H }  & \qw\\
\pureghost{   \Psi} &  \vdots &\vdots  &  \vdots &  \\
\pureghost{   \Psi} &\qw \poloFantasmaCn{\spc H}   &\gate{U_x }  & \qw \poloFantasmaCn{\spc H }  &\qw 
} }   
 \end{equation}   We refer to these strategies as \emph{ancilla-free}.  
 Note that the only difference between an ancilla-free strategy and a general parallel strategy, as in Eq. (\ref{parastrat}),  is the presence of a non-trivial ancilla system. We will now show that the ancilla system can be always traded for an additional number of queries:   
\begin{Theo}\label{theo:free}
Every  parallel discrimination strategy using a $d_A$-dimensional ancilla can be replaced by an ancilla-free strategy using a finite number of extra queries.  For large $d_A$, the minimum number of extra queries scales as $O(\log_d  (d_A))$.
\end{Theo}

This result guarantees that  a discrimination strategy using a large number  $N_A$ of ancillary qubits can be replaced by an ancilla-free strategy that uses  $O(N_A)$ extra queries to the black box. Essentially, this means that the black box queries are a stronger resource than the use ancillary qubits, as the latter can be efficiently simulated using the former.    

\subsection{Upper bounds on the query complexity of ancilla-free unambiguous discrimination}
Let us denote by $N^{AF}_{\min}$ the query complexity of ancilla-free unambiguous discrimination, i.e. the minimum number of queries needed to distinguish the gates $\set U$ unambiguously in an ancilla-free way. 
It is immediate to see that ancilla-free unambiguous discrimination is possible in  $N$ queries if and only if there exist a state $|\Psi\>  \in\spc H^{\otimes N}$ such that the output states $  |\Psi_x\>  = U_x^{\otimes N}  |\Psi\>   $ are linearly independent.
Using this fact, the upper bounds of Propositions \ref{bnd:linear} and \ref{theo:fidelity} can be easily adapted, by replacing the dimension $\dim (\set U)$ and the minimax fidelity $F_{\set U}$ with the corresponding ancilla-free quantities: 
 \begin{Proposition}\label{prop:ancfree}
 The query complexity of ancilla-free unambiguous discrimination of the gates $\set U$  is upper bounded as 
\begin{eqnarray}\label{linearanc}  N^{AF}_{\min}  \le    |\set U|  -    \dim_{loc}(\set U)  + 1   ,  
\end{eqnarray}
where  $\dim_{loc}(\set U)$ is the maximum over all possible input states $|\psi\>\in\spc H$ of the dimension of the subspace spanned by the vectors $  (  U_x |\psi\>)_{x\in\set X}$.   
Another upper bound is given by
\begin{eqnarray}\label{locfid}
N^{AF}_{\min}\le \left \lfloor  \frac{\log  (|\set U|-1)}{  \log   \left( F_{loc,\set U}^{-\frac 12}\right)}   \right \rfloor +1 \, , 
 \end{eqnarray}  
 where $   F_{loc,\set U}$ is the local minimax fidelity  $F_{loc,\set U}  :  =  \min_{|\psi\>\in\spc H}  \max_{x\not = y}    |\<  \psi | U_x^\dag U_y  |\psi\> |^2$.   
  \end{Proposition} 
The proof is the obvious  adaptation of proof for general parallel strategies, provided in the Appendix.  Note that, due to the prohibition to use ancillas, one has  $\dim_{loc} (\set U)   \le \dim(\set U)$  and   $F_{\set U,loc} \ge  F_{\set U}$. Therefore, the values of the upper bounds in Proposition \ref{prop:ancfree} are larger than the values of the upper bounds in Propositions  \ref{bnd:linear} and \ref{theo:fidelity}.  
Nevertheless,  even without the use of ancillas, the minimum number of queries needed for unambiguous discrimination is always less than  $|\set U|-1$, the minimum number of  tests that would be needed to identify a gate in $\set U$ via  pairwise eliminations.  
  The local fidelity  bound of Eq. (\ref{locfid}) provides an even better estimate when the unitaries in $\set U$ generate a SIC-POVM \cite{sic1,sic2}, in which case one has $F_{loc,\set U}  \le (d+1)^{-1}$, implying that $N= 4$  queries are sufficient for ancilla-free discrimination.  This estimate is much better than the estimate coming from the linear bound of Eq. (\ref{linearanc}), which in the case under consideration gives a quadratic scaling with the dimension $  N_{\min}^{AF}  \le   d^2-d+1$.

\subsection{Perfect ancilla-free discrimination of generalized $t$-designs }\label{subsect:groups}

For generalized $t$-designs, one can prove that unambiguous discrimination coincides with perfect discrimination, \emph{even in the case of ancilla-free strategies}: 
\begin{Proposition} Let $\set U $ be a generalized $t$-design.  Then, the following are equivalent 
\begin{enumerate}
\item there exists an ancilla-free strategy for unambiguous discrimination using $N\le t$ queries 
\item there exists an ancilla-free strategy for perfect deterministic discrimination of the  gates using $N\le t$ queries.
\end{enumerate} 
 \end{Proposition}
\medskip

A case  of special interest is  when unitaries form a group representation, namely $\set U  \equiv  (  U_g)_{g\in\grp G}$.  
Two group-theoretic  lower bounds on the query complexity  were originally derived in \cite{thesis}. We include them here for completeness, concluding our general investigation of gate identification with multiple queries:

\begin{Proposition}[\cite{thesis}]\label{bnd:delta}
For every perfect discrimination strategy using $N$ parallel queries and a $d_A$-dimensional ancilla there exists  a perfect ancilla-free strategy using $N +\Delta N_{\min}$  queries, with
\begin{equation}
\Delta N_{\min} \le   \left \lceil \frac{\log d_{A}  + \log \sqrt{|\grp G|} }{\log d} \right \rceil. 
\end{equation}
\end{Proposition}

Note that for a strategy using a large number of ancillary qubits $N_A  \gg   \log\sqrt{ |\grp G|}$, the number of extra queries to the black box scales as $O(N_A)$, as anticipated by theorem \ref{theo:free}.  Note also that the bound is independent of the number of queries of the initial ancilla-assisted strategy: in order to apply the bound, we only need to know that the original strategy allowed for perfect discrimination.    

\medskip 

The last bound is expressed in terms of the maximum entanglement fidelity between pairs of unitaries.  The entanglement fidelity between two unitaries $U_g$ and $U_h$, defined as
\begin{equation*}
F_{ent}  (  U_g,U_h)  : =  \frac{  \left|  \Tr  [  U_g^\dag U_h ]\right|^2} {d^2} \, ,
\end{equation*}
is the fidelity between the states $  |\Phi_g\>  =  (U_g \otimes I ) |\Phi\>$ and $  |\Phi_h\>  =  (U_h \otimes I ) |\Phi\>$, obtained by applying the two unitaries on one side of the maximally entangled state $  |\Phi\>  =  (\sum_{n=1}^d  |n\>  |n\>)/\sqrt{d}$.    
Thanks to the group structure, the maximum entanglement fidelity over all pairs of unitaries in $\set U$ is given by
$F_{ent,\set U}   =  \max_{g\in\grp G}  {\left|  \Tr[  U_g]\right|^2}/{d^2}$.     

The other quantity appearing in the bound is the number of unitaries $U_g$ that can be confused with the identity, given by
\begin{equation*}
C  =   \left|   \{  U_g\not = I  ~|~    F_{ent}  (U_g,I) \not = 0  \} \right|  \, .
\end{equation*}

With the above definitions we have the following
\begin{Proposition}[Entanglement fidelity bound \cite{thesis}]\labell{Prop:ConditionAlpha}
 If  
\begin{equation}\label{eeee}
{d^N }~ \left( 1-   F_{ent,\set U}^{N/2} ~C  ~\right)  \ge |\grp G|,
\end{equation} 
then the unitaries in $\set U$ can be perfectly distinguished with an ancilla-free strategy using  $N$ queries.
 \end{Proposition} 
 
 \medskip  
The condition of Eq. (\ref{eeee})  contains implicitly an upper bound on the query complexity of ancilla-free discrimination. The fact that an \emph{upper bound} on $N^{AF}_{\min}$ can be expressed in terms of the entanglement fidelity is quite surprising, because applying the unitaries on one side of a maximally entangled state is not an allowed strategy in ancilla-free gate discrimination.  

Let us  discuss the consequences of Eq. (\ref{eeee}).  The most immediate consequence is that when the entanglement fidelity is zero, perfect ancilla-free discrimination is possible with $N  =  \lceil  \log_d |\grp G|  \rceil$ queries.   
This value is actually the optimal one, because  $\lceil  \log_d |\grp G| \rceil$ is the minimum number of copies that is needed to pack $|\grp G|$ orthogonal vectors in the $N$-fold tensor product  of a $d$-dimensional Hilbert space.
An example of this situation is the discrimination of shift-and-multiply gates, already discussed in Example \ref{ex:shiftmult}:  
\begin{Example}[Ancilla-free discrimination of shift-and-multiply gates]  
For a $d$-dimensional quantum system, the shift-and-multiply gates $U_{pq}  =  S^p  M^q$ are $d^2$ mutually orthogonal unitaries.   Eq. (\ref{eeee}) then predicts that perfect ancilla-free discrimination  is possible with $N^{AF}_{\min} = 2$ queries.  A concrete discrimination strategy consists in preparing two probe systems in the state  $|0\>  |f_0\>$, where $|f_0\> =  \left( \sum_{n=1}^d    |n\>   \right)/\sqrt{d}$ is the first vector of the Fourier basis.   Applying two queries of the unknown gate $U_{pq}$ we then obtain the output state $|p\> |q\>$, from which $p$ and $q$ can be read out in a perfect deterministic way. 
\end{Example}

Before concluding, we note that,  more generally, the optimal scaling $N^{AF}_{\min} =  O(\log_d  |\grp G|)$  
can be achieved whenever the condition
\begin{equation}\label{bellissima}
\frac{\log \left(  \frac{C }{ 1-\alpha } \right)  } {\log \left( F_{ent,\set U}^{-\frac 1 2} \right) } \le \log_d  |\grp G|
\end{equation} 
is satisfied for some constant $\alpha  >  0$.  Indeed, under this condition the query complexity of ancilla-free gate discrimination can be bounded between  $ \lceil \log_d |\grp G| \rceil $ and    $\lceil \log_d |\grp G| \rceil + \log_d  \alpha^{-1}$, thus implying 
\begin{equation*}
N_{\min}^{AF}  =  O(  \log_d |\grp G|). 
\end{equation*} The argument is simple and is provided in the Appendix.

\section{Conclusion}\label{sec:conclusions}
 
In this paper we  investigated the problem of identifying  an unknown unitary gate in a finite set of alternatives, using  both deterministic and probabilistic discrimination strategies, and allowing the unknown gate to  be queried multiple times and to be be used in parallel or in series in  arbitrary quantum circuits.   In this scenario, we provided upper and lower bounds on the amount of resources needed to achieve unambiguous and perfect gate identification.  Specifically, we gave bounds on  the query complexity and the minimum size of the ancillas. 

Most of our results stem from two key observations.    The first observation is that unambiguous gate discrimination can be parallelized: if unambiguous discrimination is possible with $N$ queries, then unambiguous gate discrimination must also be possible by applying the $N$ queries in parallel on a suitable entangled state.     
The second key observation is based on the definition of generalized $t$-designs, a definition that includes unitary $t$-designs and group representations as special cases.    
The remarkable feature of  generalized $t$-designs is that for strategies using $N \le t$ queries there is no difference between unambiguous and perfect discrimination. 
Using this fact, one can reduce the analysis of perfect gate discrimination to the simpler analysis of unambiguous gate discrimination.   
Finally, motivated by the application to quantum communication in the lack of shared reference frames, we considered discrimination strategies where ancillas are not allowed,  providing upper bounds on the number of extra queries that are needed to make up for this limitation.   

Our results suggest several directions of further research. First of all, up to now  we considered the question whether or not  unambiguous discrimination is possible with a certain number of queries. However, as the example of Grover's algorithm clearly shows, sometimes the probability of unambiguous discrimination could be too small to have a useful application.   Hence, for future developments it is important to have bounds on the probability of the inconclusive outcome as a function of the number of queries.  Our work addressed the question in the simplest case, namely the case of generalized $t$-designs, where  the probability of the inconclusive outcome is always zero.  In other cases, like the discrimination of Grover's unitaries, an estimate of the probability of the inconclusive outcome as a function of the number of queries can be obtained with the techniques developed by Eldar \cite{eldar}. Our work provides a useful first step in this direction, because knowing that the unitaries are unambiguously distinguishable, and therefore, that they produce linearly independent output states is  a necessary condition for the application of the techniques of Ref. \cite{eldar}.  Besides the evaluation of the probability of the inconclusive outcome, it is also worth relaxing the requirement of perfect gate identification, allowing for a small probability of error  $p_e \le \epsilon$. In this case, the interesting quantity would be the minimum number of queries $N_{\min,\epsilon}$ needed to achieve  gate identification with error probability smaller than $\epsilon$. For generalized $t$-designs, our expression for the maximum probability of correct discrimination, given by $p_{N}^{\max}  =   \dim (\set U_N)/|\set U |$ (cf. theorem \ref{theo:design}), gives  a starting point for the evaluation of $N_{\min,\epsilon}$. 

Another interesting development suggested by our work is the experimental demonstration of quantum advantages in  gate discrimination with multiple queries.   While the viability of unambiguous state discrimination has been demonstrated in many experiments (see e.g. \cite{exp1,exp2,exp3,exp4}), the experimental realization of gate discrimination strategies is a rather unexplored territory. A recent experiment demonstrating unambiguous discrimination of two non-orthogonal gates was reported in Ref. \cite{obrian} in the single-query scenario. As the level of control in the experiments increases, it would be highly desirable to have proof of principle demonstrations of the advantage of joint multi-query discrimination strategies over  strategies based on pairwise elimination, both in the case of perfect discrimination and of unambiguous gate discrimination.   Moreover, our results on gate discrimination without ancillas show suggest one may have quantum advantages even with relatively small amounts of entanglement.        

\bigskip 
{\bf Acknowledgments. }   
This work was carried out while MR was with NEC Laboratories America, Princeton, NJ 08540, USA.
GC is supported by the National Basic Research Program of China (973) 2011CBA00300 (2011CBA00301), by the National Natural Science Foundation of China (Grants 11350110207, 61033001,  61061130540), and by the 1000 Youth Fellowship Program of China.  GC acknowledges helpful discussions with H Zhu, M Graydon, and HB Dang on the linear independence of SIC-POVM states and the hospitality of Perimeter Institute, where part of this research was done. Research at Perimeter Institute for Theoretical Physics is supported in part by the Government of Canada through NSERC and by the Province of Ontario through MRI. 

\section*{Appendix}

\subsection*{Proof of theorem \ref{theo:unambiguous}.} 

We first prove necessity. The condition for error-free discrimination is equivalent to the existence of at least one $x_0 \in\set X$
 such that $p_N(  x_0 |  x) =  0 \quad \forall x\not =  x_0$, which in turn is equivalent to the condition that the output state $|\Psi_{x_0}\>$ in Eq. (\ref{outstate}) is linearly independent from the states $ (|\Psi_x\>)_{x\not =  x_0}$.   Since the function $ U_x^{\otimes N}  \mapsto   |\Psi_x\>$ is linear, the condition $U^{\otimes N}_{x_0}   \not \in  \Span  (  U^{\otimes N}_{x})_{ x\not = x_0}$ is necessary for error-free discrimination.     Similarly, the condition for unambiguous discrimination is equivalent to requirement that $p_N(x_0| x)  =   0 \quad \forall  x\not = x_0$, which in turn is equivalent to the requirement that the output states  $( |\Psi_x\>)_{x\in\set X}$ are linearly independent.  Independence of the states $(|\Psi_x\>)_{x\in\set X}$ implies independence of the unitaries $(U_x^{\otimes N})_{x\in\set X}$.  Both conditions are also sufficient, because  the linear function  $   U^{\otimes N}_x \mapsto   |\Phi_x\>^{\otimes N} $ defined by $|\Phi_x\>:  =  (U_x \otimes I) |\Phi\>$,  $|\Phi\>  :   =  \sum_{n=1}^d  |n\>|n\>/\sqrt{d}$ is invertible, and therefore preserves linear independence.      Note that the states $|\Phi_x\>$ can be obtained from a parallel strategy where  $N$ pairs of systems are prepared in the state $|\Phi\>^{\otimes N}$ and the unitary $U_x$ is applied on the first system of each pair. \qed 

\subsection*{Proof of proposition 1}
By theorem \ref{theo:unambiguous}, unambiguous discrimination is possible only if $\dim  (\set U_N )  =  |\set U|$.       On the other hand,  thinking of each unitary $ U^{\otimes N} $ as a vector in the symmetric subspace of $\left(\Cmplx^{d^2}  \right)^{\otimes N}$ we have $\dim \set U_{N} \le\left(   \begin{array}{c}
   N + d^2-1 \\
 d^2-1 
 \end{array}
 \right) $.    The bound is tight, because one can always choose the unitaries $\set U_N$ to be a spanning set for  the symmetric subspace of $\left(\Cmplx^{d^2}  \right)^{\otimes N}$.
\qed

\subsection*{Proof of proposition \ref{bnd:linear}}

 Let  $\set S  =(v_x)_{x\in \set X}$ be a finite set of vectors in a vector space $V$, with the property that every two distinct vectors in $\set S$ are linearly independent.   Under this hypothesis, Chefles  proved that $\dim \Span (v_x^{\otimes N+1})  \ge \dim \Span (v_x^{\otimes N}) +1$ \cite{Chefles}.    Applying the result to the set $\set U_N  :=(U_x^{\otimes N})_{x\in\set X}$ gives $\dim (\set U_N) \ge \dim ( \set U ) +   N-1$.   Hence, for the unitaries in $\set U_N$ are linearly independent for $N   = |  \set U|   -  \dim  (\set U)  +1 $.  \qed  
 
\subsection*{Proof of lemma 2} 

 Suppose that $\sum_{y\in \set X}   c_y   |\psi_y\>^{\otimes N}  =0$.  Multiplying by $\< \psi_x|^{\otimes N}$, taking the modulus, and summing over $x$ we obtain  
\begin{eqnarray*}
\sum_{x\in \set X}|c_x|   &  =      \sum_{x\in\set X} \left|\sum_{y\in \set X, y \not = x}   c_y  \<  \psi_x  |  \psi_y\>^N  \right|  \\
 &  \le     \sum_{x\in\set X} \sum_{y \in\set X, y \not = x}  |c_y|  F^{N/2}\\
   &  =   (|\set X|  -1)  F^{N/2}      \left(  \sum_{x\in \set X}|c_x|\right) . 
\end{eqnarray*}   
Clearly, if   $(|\set X|  -1)  F_{\set S}^{N/2}  <  1$, the only possible solution is $c_x  =  0~  \forall x\in\set X$. Hence, the states  $(  |\psi_x\>^{\otimes N})_{x\in\set X}$ are linearly independent.      \qed 

An alternative proof of lemma 2 can be obtained from the application of the Welch bound \cite{welch}. 
\medskip 

\subsection*{Proof of proposition 3} 
Choose the input state $|\Psi\>\in\spc H\otimes \spc H$ so that  $\max_{x,y\in\set X,x\not = y}  |\<  \Psi  | (U_x^\dag U_y\otimes I) |\Psi\>|^2  = F_{\set U}$.  For $F_{\set U}^{N/2}  \le  1/(|\set U|-1) $ the states $(|\Psi_x\>^{\otimes N} )_{x\in\set X}, |\Psi_x\>  :  =  (U_x \otimes I)|\Psi\>$ are linearly independent. Therefore, also the unitaries $(U_x^{\otimes N})_{x\in \set X}$ are linearly independent, i.e. unambiguously distinguishable. \qed

\subsection*{Proof of theorem 2}
 Using the formalism of quantum combs \cite{combprl,comblong}, we express  the probability $p_N(y|x)$ as $ p_N(y|x)  =    \<\!\<  U_x  |^{\otimes N}  ~    T_y ~  |U_x\>\!\>^{\otimes N}$ where   $ ( T_y)_{y\in\set Y}$ is a collection of positive operators satisfying suitable normalization conditions \cite{watgut,combprl} (the actual form of the conditions is irrelevant here). 
The probability of correct identification can be bounded as
\begin{eqnarray*}
p_{N}  &   =  \frac {\sum_{x\in\set X}   p_x ~     \<\!\<  U_x  |^{\otimes N}  R_N^{-\frac 12}~     \left (  R_N^{\frac 12} T_x   R_N^{\frac 12}  \right) ~   R_N^{-\frac 12}  |U_x\>\!\>^{\otimes N}  } {  \sum_{  y  \in \set X}    \Tr [  T_y    R_N]} \\
  &  =  \sum_{x\in\set X}         p_x ~  \Tr  [    \rho_x   ~   R_N^{-\frac 12}    (   |U_x\>\!\>   \<\!\<  U_x  |)^{\otimes N}  R_N^{-\frac 12} ]    \qquad    \rho_x  :  =    \frac{   R_N^{\frac 12}    T_x  R_N^{\frac 12} }  {   \sum_{y\in\set X}   \Tr [R_N^{\frac 12}    T_y  R_N^{\frac 12} ]} \\
    &\le   \sum_{x\in\set X}         p_x ~  \Tr  [    \rho_x ]     ~ \|    R_N^{-\frac 12}    (   |U_x\>\!\>   \<\!\<  U_x  |)^{\otimes N}  R_N^{-\frac 12} \|_{\infty}  \\  & \le \max_{x\in\set X}  p_x ~   \<\!\<  U_x  |^{\otimes N} ~  R_N^{-1} ~  |U_x\>\!\>^{\otimes N} ,     \end{eqnarray*}
the last inequality coming from the condition $\sum_{x\in\set X}  \Tr[\rho_x]  =1$.      Defining 
\begin{eqnarray*}
 x_{\max}  :  = {\rm argmax} ~  p_x ~   \<\!\<  U_x  |^{\otimes N} ~  R^{-1} ~  |U_x\>\!\>^{\otimes N} ,
 \end{eqnarray*} the bound can be saturated by applying the $N$ queries of $U_x$ in parallel on the maximally entangled state $|\Phi\>^{\otimes N} $,$  |\Phi\>  :  =  |I\>\!\> /\sqrt d$, and by performing the POVM  $(P_y)_{y\in\set  Y}$ defined by $P_{x_{\max}}   =   R^{-1}    (|  U_{x_{\max}}\>\!\>   \<\!\<  U_{x_{\max}}  |  )^{\otimes N}   R^{-1}  /   \<\!\<  U_{x_{\max}}  |  )^{\otimes N}    R^{-2} |  U_{x_{\max}}\>\!\> $,  $  P_? =   I  -   P_{x_{\max}}$, $P_y  = 0$ for every $y  \not  =  x_{\max}$. \qed   
 
\medskip

\subsection*{Basic representation-theoretic facts}\label{subsec:designsrep}

Since generalized $t$-designs have an underlying group-theoretic structure, before proceeding to the next proofs it is useful to  recall  some basic facts about group representations.   

Let us denote by $\Lin (\spc H)$ the set of linear operators acting on $\spc H$ and consider a  representation  $U:    \grp G  \to    \Lin(\spc H),   ~g \mapsto U_g $ of some (compact) group $\grp G$.  Here we allow $U$ to   be a    a \emph{unitary projective representation (UPR)}, with multiplier  $\omega: \grp G \times \grp G  \to \Cmplx$. In short, this means that $U_g U_h  =   \omega (g,h)   U_{gh}, ~\forall g,h\in\grp G$.   The most familiar case is the case of the unitary representations (UR), for which $\omega (g,h) =1 ~\forall g,h\in\grp G$.  

With a suitable choice of basis, the Hilbert space can be decomposed as  a direct sum of tensor product pairs
\begin{equation}\label{cgh}
\spc H = \bigoplus_{\mu \in\Irr (U)}   \left(   \spc R_\mu  \otimes \spc M_\mu \right),
\end{equation} 
where the sum runs over the set $\Irr (U)$ of all inequivalent  irreducible representations (\emph{irreps})  contained in the decomposition of $U$, $\spc R_\mu$ is a \emph{representation space}  of dimension $d_\mu$, carrying the irrep $U^\mu$, and $\spc M_\mu$ is a \emph{multiplicity space} of dimension $m_\mu$, $m_\mu$ being the multiplicity of the irrep $U^\mu$ in the decomposition of $U$.  Eq. (\ref{cgh}) implies that the representation $U$ can be written in the block diagonal form
\begin{equation}\label{cg}
U =  \bigoplus_{\mu \in \Irr (U)}  \left( U^{\mu}  \otimes I_{\spc M_\mu}  \right) ,
\end{equation} 
where $I_{\spc M_\mu}$ denotes the identity matrix on  $\spc M_\mu$.    Note that all the irreps $U^\mu \in \Irr  (U)$ must have the same multiplier $\omega$.  

Using Eq. (\ref{cg}) and the orthogonality of matrix elements, one can prove that  the set of unitaries  $\set U   :=  ( U_g)_{g\in\grp G}$  satisfies  
\begin{eqnarray}\label{dimension}
\dim (\set U)   =   \sum_{\mu  \in\Irr (U)}   d_\mu^2 .  
\end{eqnarray} 
Due to the importance of linear independence in the gate discrimination problem, this equation will become very useful in the following section.  

A representation that plays a key role in gate discrimination is  the \emph{regular representation}, which for finite groups is a representation of $\grp G$ on the Hilbert space  $\spc H  = \Cmplx^{|\grp G|}$, equipped  with the orthonormal basis $\{|g\>~|~ g \in \grp
G\}$. 
Precisely, the regular representation with multiplier $\omega$  is the  projective representation $U^{reg,\omega}  :      \grp G  \to \Lin (\Cmplx^{|\grp G|})$ defined by 
\begin{equation}\label{Def:RegRep}
U^{reg,\omega}_g |h\> = \omega(g,h)~|gh\>~, \qquad \forall g,h  \in   \grp G
\end{equation}

\medskip 




The regular decomposition is reducible and  its  decomposition  is 
\begin{equation}\labell{RegRepUnitDecomp}
U_g^{reg,\omega} = \bigoplus_{\mu \in {\rm Irr}(\grp G, \omega)} ~  \left(   U_g^{\mu} \otimes I_{\spc M_\mu}  \right) \qquad \spc M_\mu  \simeq \Cmplx^{d_\mu}
\end{equation}
where ${\rm Irr}(\grp G, \omega)$ denotes the set of all the irreps
 of $\grp G$ with  multiplier $\omega$ [in particular, ${\rm Irr}(\grp G, 1)$ is the set of all \emph{unitary} irreps of $\grp G$].     Note that every irrep appears with multiplicity $m_\mu  =  d_\mu$.  Choosing $g = e$ (the identity element in the group) and  taking the trace on both sides of Eq. (\ref{RegRepUnitDecomp})   one  obtains 
\begin{equation}\labell{FiniteGroupCard}
|\grp G| = \sum_{\mu \in {\rm Irr}(\grp G, \omega)} d_{\mu}^2~, 
\end{equation}
which holds for every possible multiplier $\omega$.  Finally, combining Eqs. (\ref{dimension}) and (\ref{FiniteGroupCard}), one gets the following statement:
\begin{Proposition}\label{prop:groups}
Let $\grp G$ be a finite group and let $U  :      \grp G  \to \Lin (\spc H)$ be a UPR with multiplier $\omega$.  Then, the unitaries $(U_g)_{g\in\grp G}$ are linearly independent if and only if the decomposition of $U$ contains all the irreps in ${\Irr}  (\grp G,\omega)$.
\end{Proposition}

\subsection*{Proof of theorem \ref{theo:design}}
Let $\grp G$ the compact group such that  $\sum_{x\in\set X}   (U_x \otimes \overline {U}_x)^{\otimes N}   =    \int {\rm d}g ~    (U_g \otimes \overline {U}_g)^{\otimes N} $, or equivalently, $\sum_{x\in\set X}   U_x^{\otimes N}  A     U_x^{\dag\otimes N} =  \int {\rm d}g ~  U_g^{\otimes N}  A     U_g^{\dag\otimes N} $ for every operator $A\in\Lin (\spc H^{\otimes N})$.  Exploiting the  decomposition of $U^{\otimes N}$, one can write $U_x^{\otimes N}  =  \bigoplus_{\mu\in\Irr (U)}  \left ( U^{\mu}_{x} \otimes I_{\spc M_\mu}\right) $ and, therefore,   $ |U_x\>\!\>^{\otimes N}  =  \bigoplus_{\mu\in\Irr(U^{\otimes N})}   |U_x^{\mu}\>\!\>  |I_{\spc M_\mu}\>\!\>$.    The operator $R_N$ in theorem \ref{theo:optprob} can be directly computed as  
\begin{eqnarray*}
R_N    & = \sum_{x\in \set X}   p_x    ~  (  |U_x\>\!\>\<\!\<  U_x |  )^{\otimes N}  \\  
  &  =  \int {\rm d}g ~   (  |U_g\>\!\>\<\!\<  U_g |  )^{\otimes N} \\
    & =   \bigoplus_{\mu\in \Irr (U^{\otimes N})}    \frac {m_\mu}{d_\mu}  ~  \left(    I_{\spc R_\mu}  \otimes  I_{\spc R_\mu}   
    \otimes     \frac{  |I_{\spc M_\mu}  \>\!\>\<\!\<   I_{\spc M_{\mu}}| }{m_{\mu}} \right), 
\end{eqnarray*}
so that, computing the inverse, one has $  \<\!\<  U_x  |^{\otimes N} ~  R_N^{-1} ~  |U_x\>\!\>^{\otimes N}    =  \sum_{\mu \in\Irr (U^{\otimes N})}  d_\mu^2   =  \dim  (\set U^{\otimes N})$  [cf. Eq. (\ref{dimension})].   Inserting this value in Eq. (\ref{pmax}) proves Eq. (\ref{pmaxdesign}).   We now prove that for the uniform prior the maximum success probability can be obtained with a deterministic strategy that uses the $N$ queries in parallel.    To this purpose, consider the maximum likelihood input state \cite{covlik,degiorgi}:  this is  the state in $\spc H^{\otimes N} \otimes \spc H_A$ given by 
\begin{eqnarray*}
|\Phi_{ML}\>:  =  \bigoplus_{\mu \in \Irr (U^{\otimes N})}   \sqrt{\frac{d_\mu}{\dim (U^{\otimes N})} }~   |I_{\spc R_\mu}\>\!\>,
\end{eqnarray*}     
where  $|I_{\spc R_{\mu}}\>\!\>  =  \sum_{n  = 1}^{d_\mu}   |\alpha^{\mu}_n\> |\beta^{\mu}_n\>$, $(|\alpha^{\mu}_n\>)_{n=1}^{d_\mu} $ being an orthonormal basis for $\spc R_\mu$ and $(|\beta^{\mu}_n\>)_{n=1}^{d_\mu} $  being an orthonormal set of vectors in $\spc  M_\mu \otimes \spc H_A$ [here the dimension of $\spc H_A$ is chosen in order to satisfy the relation $d_{\mu}  \le   m_{\mu}   d_A , \forall \mu\in\Irr (U^{\otimes N})$].      Applying the $N$ queries in parallel one obtains the output states $  |\Phi_{ML,x}\>  : =  (U_x^{\otimes  N} \otimes I_A)  |\Phi_{ML}\>$. Optimal discrimination can be achieved deterministically using  the square root measurement \cite{squareroot}, which in this case has POVM elements $  P_x  :=   \frac{\dim (\set U_N)}{  |\set U|} ~ |\Phi_{ML,x}\>\<  \Phi_{ML,x} |$.      \qed

\subsection*{Proof of proposition \ref{bnd:ancilla}}

The proof is an immediate generalization of the proof of Lemma 1 in Ref. \cite{entest}.  We provide it here just for the sake of completeness.  Consider the irreducible  decomposition $\spc H^{\otimes N}  =  \bigoplus_{\mu \in  {\rm Irr}    (U^{\otimes N})}    \spc R_\mu  \otimes \spc M_{\mu}$ associated to the group representation $U: g  \mapsto U_g$.    Take an ancillary Hilbert space  $\spc H_A$  of dimension   $ d_A  = \max_{\mu  \in  {\rm Irr}  (U^{\otimes N})} \left\lceil {d_\mu}/{m_\mu} \right\rceil $ and define  $  |\Phi\>  \in  \spc H^{\otimes N}  \otimes \spc H_A $ to be the unit vector $  |\Phi\>  =  \bigoplus_{\mu\in  {\rm Irr} (  U^{\otimes N})}    c_\mu   |\Phi_{\mu}  \> $, where   $c_{\mu}$ are non-zero coefficients and $|\Phi_{\mu}\>$ is a maximally entangled state in   $\spc R_{\mu}   \otimes   \spc M_{\mu}'$, $\spc M_\mu'  : =  \spc M_\mu \otimes \spc H_{A}$.      Then, the number of linearly independent states of the form $ |\Phi_x\>  :  =    (U_x^{\otimes N}  \otimes I_A)  |\Phi\> , {x\in\set X}$ is equal to the number of linearly independent unitaries in the set $(U^{\otimes N}_x)_{x\in\set X}$.  In particular,  the unitaries  $(U^{\otimes N}_x)_{x\in\set X}$ are linearly independent (i.e. unambiguously distinguishable) iff the states  $(|\Phi_x\>)_{x\in\set X}$  are linearly independent (i.e. unambiguously distinguishable), that is, iff unambiguous discrimination is possible using a $d_A$-dimensional ancilla satisfying  $ d_A  = \max_{\mu  \in  {\rm Irr}  (U^{\otimes N})} \left\lceil {d_\mu}/{m_\mu} \right\rceil $.  \qed

\subsection*{Proof of corollary 4} 

Every unitary set $\set U$ is contained in the group $U(d)$.  The irreps of the tensor representation $U \mapsto U^{\otimes N}$ are labelled by Young diagrams of $N$ boxes arranged in at most $d$ rows and their dimensions (multiplicities) are given by  \cite{fultonharris}
\begin{eqnarray}\label{hook}
d_{\mu}   =  \prod_{(i,j)  \in   \mu}      \frac{  d  +  j-i}{  |  h_{ij}| }   \qquad \left ( m_{\mu}   =      \frac{  N!}{    \prod_{(i,j)  \in   \mu}  |  h_{ij}| }  \right),
\end{eqnarray}
where the products runs over the boxes in the Young diagram $\mu$, each box being identified by its row-column coordinates $(i,j)$.   Here, $|h_{ij}|$ is the length of the hook centred on the box $(i,j)$.   Taking the ratio, one obtains 
\begin{equation*}
\frac {d_{\mu}}{m_\mu} \le       \frac{  \prod_{(i,j)  \in   \mu}    d  +  j-i}{ N!}   \le   \left(     \begin{array}{c}  N  +  d  -1  \\  d-1  \end{array}  \right)   \qquad \forall \mu  \in {\rm Irr}  (U^{\otimes N}). 
\end{equation*}
\qed

\subsection*{Proof of theorem 4}

The idea is to encode the ancilla space into one of the multiplicity spaces $\spc M_{\mu_0}$ contained in the decomposition of the $M$-fold tensor representation of $U(d)$. This can be done using the invariant encoding of Ref. \cite{prlcommun}.    Precisely, choosing  $\mu_0\in \Irr (U^{\otimes M})$ such that  $m_{\mu_0} \ge d_A$, one can encode the input state $|\Psi\>\in  \spc H^{\otimes N}  \otimes \spc H_A$ through the isometric embedding $V_{\mu_0} :  \spc H^{\otimes N} \otimes \spc H_A    \to \spc H^{\otimes  N}  \otimes \spc R_{\mu_0}   \otimes \spc M_{\mu_0}   \subset \spc H^{\otimes (N + M)}$ defined by $V_{\mu_0}   |\alpha\>  |\beta\>  :=  |\alpha\>  |\psi_0\>|\beta\>$, $|\psi_0\> \in \spc  R_\mu$ being a fixed unit vector.  One way to satisfy the condition   $m_{\mu_0} \ge d_A$ is to choose $  M  =  d \cdot l$ and to set $\mu_0$ to be the Young diagram with $d$ rows of length $l$.    
By Eq. (\ref{hook}) one has 
\begin{equation*}
m_{\mu_0}  =     \frac  {  M!}{  ( l!  )^d    \prod_{k=0}^{d-1}    \left (  \begin{array}{c} l  + k \\k \end{array}\right)   }    \ge \frac{ M!}{ (l!)^d   \prod_{k=0}^{d-1}  (l+1)^k}    =   \frac{ M!}{ (l!)^d   (l+1)^{d(d-1)/2}}.
\end{equation*} 
Hence, for large $M$ the Stirling approximation yields $\log_d  (m_{\mu_0})  =    M  \log_d  (M/e)   -  M \log_d  (l/e)   -  O  (\log_d  l)      =  M  -  O(\log M) $.  For large $d_A$, the condition $ m_{\mu_0}\ge d_A$ is then satisfied whenever $  M  \ge  (1+  \epsilon) \log_d  (d_A) $, $\epsilon > 0$.   
\qed

\subsection*{Proof of proposition 6}

Suppose that the states  $|\Psi_x\>  := U_x^{\otimes N}  |\Psi\>   $ are linearly independent and define $\rho  :  = 1/|\set U|  \sum_{x\in\set X}    |\Psi_x\>\< \Psi_x|$.    Then, the states  $|\Phi_x\>  : = \sqrt{1/|\set U|} \rho^{-1/2}|\Psi_x\>$ are mutually orthogonal.    
  Since the unitaries form an $t$-design, we have $ [\rho,  U_x^{\otimes N}] = 0 \quad \forall x\in\set X$. Hence, the orthogonal states $|\Phi_x\>$ are generated by applying  $U_x^{\otimes N}$ to the state $|\Phi\>  :  = \sqrt{1/|\set U|} \rho^{-1/2}|\Psi\>$.  This yields the desired strategies for perfect discrimination. \qed

\subsection*{Proof of proposition \ref{bnd:delta}} 
The proof invokes the following bound on the sum of the multiplicities arising in a tensor representation:
\begin{Lemma}\labell{Lemma:TotalMult}
Let $m_\mu$ be the multiplicity of the irrep $\mu$ in the decomposition of $U^{\otimes M}$.  Then, 
\begin{equation}\labell{TotalMult}
\sum_{\mu \in \Irr (U^{\otimes M})} m_{\mu} \ge \frac{ d^M}{\sqrt{|\grp G|}}  \end{equation}
\end{Lemma}
\Proof    Setting $g=e$ in the  decomposition $U_g^{\otimes M} = \bigoplus_{\mu \in \Irr (U^{\otimes M)}}   \left(  U_g^{\mu} \otimes I_{\spc M_{\mu}} \right)$, and taking the trace of $U_g^{\otimes M}$ one gets $d^M  =  \sum_{\mu \in {\rm Irr}  (U^{\otimes M})  } ~d_{\mu} m_{\mu}$. On the other hand,    $d_{\mu} \le \sqrt{|\grp G|}$ for any $\mu$, whence Eq. (\ref{TotalMult}). \qed

\medskip

{\bf Proof of proposition \ref{bnd:delta}.} 
Since $N$  queries are sufficient for perfect discrimination using an ancilla, the unitaries $(U_g^{\otimes N})_{g\in\grp G}$ are linearly independent (lemma \ref{theo:unambiguous}).     Now, suppose that $M$  additional queries are available and consider the  decomposition  $U^{\otimes M} = \bigoplus_{\mu \in \Irr (U^{\otimes M)}}   \left(  U_{\nu} \otimes I_{\spc M_{\nu}} \right)$.       
 Since the tensor product cannot decrease the number of linearly independent vectors,
 the unitaries $(U_g^{\otimes N}
\otimes U_g^{\mu})_{g\in\grp G}$ are linearly independent for every $\mu\in  {\Irr}  (U^{\otimes M})$.  Equivalently, this means that the representation $U^{\otimes N}  \otimes U^\mu$ contains all the irreps with multiplier $\omega^{N+ M}$, namely $\Irr (U^{\otimes N }\otimes U^{\mu})  \equiv  \Irr (\grp G,\omega^{N+M})\equiv  \Irr (U^{\otimes (N + M)})$. 
 Denoting by $m_{\nu,  N+ M} $ the multiplicity of the representation $\nu \in  {\Irr}  (U^{\otimes N + M})$ and by $\spc M_{\nu, N+ M}$ the corresponding multiplicity space,  we have 
  \begin{eqnarray*}
  m_{\nu,N + M}  \ge   \sum_{\mu  \in {\rm Irr}  (U^{\otimes M})}  m_{\mu}   \ge \frac {d^M}{\sqrt {|\grp G|}},
  \end{eqnarray*}
  the last inequality due to lemma \ref{Lemma:TotalMult}.  
Now, the condition $d^M \ge   d_{A}\sqrt{|\grp G|}$ guarantees   
 $ m_{\nu, N+  M}  \ge    d_{A} $ 
  for every possible irrep $\nu  \in \Irr (\grp G,  \omega^{M+N})$.   Hence,  by proposition \ref{bnd:ancilla} the optimal discrimination can be achieved without ancilla.    Since $\dim ( \set U_N  )  =   |\grp G|  =  \dim (  \set U_{ N+ M})$, the discrimination is perfect (theorem \ref{theo:design}).   \qed

 \subsection*{Proof of proposition \ref{Prop:ConditionAlpha} }  
 
The proof takes advantage of the connection between perfect distinguishability and the regular representation:

\begin{Lemma}\label{prop:reg} \cite{thesis,skoti} Let  $U : \grp G  \to \Lin (\spc H)$ be a  UPR with multiplier $\omega$.   Then the following are equivalent  
\begin{enumerate} 
\item the gates  $(U^{\otimes N}_g)_{ g\in\grp G}$  are perfectly/unambiguously distinguishable without ancilla
\item  $  U^{\otimes N}$ contains as a sub-representation the regular representation with multiplier $\omega^N$,  defined by  
\begin{equation*}
U^{reg}_g |h\> = \omega^N(g,h)~|gh\>~, \qquad \forall g,h  \in   \grp G,
\end{equation*} 
where $(|g\>)_{g\in\set G}$ are orthonormal vectors.
\item  the decomposition of $  U^{\otimes N}$ contains every irrep $\mu \in\Irr (\grp G,\omega^N)$ with multiplicity $m_\mu  \ge d_\mu$. 
\end{enumerate}
\end{Lemma} 

\medskip 

{\bf Proof of proposition \ref{Prop:ConditionAlpha} }.  Let $U^{\otimes N}  =  \bigoplus_{\mu\in\Irr (U^{\otimes N})}   U^\mu\otimes I_{\spc M_{\mu}}$ be the decomposition of $U^{\otimes N}$. By the orthogonality of the characters, the multiplicity $m_{\mu}$ is given by 
\begin{eqnarray*}
m_{\mu}  =   \frac 1 {|\grp G|}  \sum_{g\in\grp G}   \overline{\Tr[U_g^\mu]}~   \Tr  [  U_g]^N  . 
\end{eqnarray*}
Hence, defining the normalized characters $  \nu (g) : =\Tr [U_g]/d $ and $\nu_\mu(g)  :=  \Tr[ U^\mu_g ]/d_\mu$,  one has
\begin{eqnarray*}
m_{\mu}   &=   \frac {d_\mu  d^N} {|\grp G|}  \sum_{g\in\grp G}   \overline{\nu_\mu  (g)}~  \nu^N(g)  \\
&\ge     \frac {d_\mu  d^N} {|\grp G|}       (1 - \sum_{g\in\Supp(\nu)\setminus \{e\} }    | \nu_\mu  (g)~  \nu^N(g)|  )\\
&\ge     \frac {d_\mu  d^N} {|\grp G|}       (1  -    F_{\set U,ent}^{N/2}  ~   C ).
\end{eqnarray*}  
If $N$ is such that  $  d^N      (1  -   F_{\set U,ent}^{N/2} ~ C ) \ge |\grp G| $, then we have $m_{\mu}  \ge d_\mu$ for every $\mu \in  {\rm Irr}  (\grp G, \omega^N)$.  This means that $U^{\otimes N}$ contains the regular representation $U^{reg,\omega^N}$ as a subrepresentation, and, therefore, perfect discrimination is possible by proposition \ref{prop:reg}. \qed 

\subsection{Scaling of the ancilla-free query complexity under the condition of Eq. (\ref{bellissima})}

Eq. (\ref{bellissima}) can be  rewritten as  
\begin{equation*}
 \left(      \frac1{  F_{ent,\set U}^{1/2} } \right)^{\log_d  |\grp G|}   \ge   \frac{C}{1-\alpha}  \, ,
\end{equation*}  
or, equivalently,  $     F_{ent, \set U}^{ \log_d |\grp G| /2}  C  \le  1-\alpha$.   
By monotonicity of the exponential, this implies that, for every $N \ge   \log_d  |\grp G|$, one has $F_{loc,\set U}^{N/2} ~  C \le   1-\alpha$, which in turn implies 
\begin{equation*}
d^N  \left(  1-  F_{loc,\set U}^{N/2}  ~C\right)  \ge  \alpha~ d^{N}.
\end{equation*}
 Hence, choosing $N  =   \lceil \log_d |\grp G| \rceil + \log_d  \alpha^{-1}$ the condition of Eq. (\ref{eeee}) is satisfied.  In conclusion,  the query complexity has been bounded  has $N_{\min}^{AF}  \le    \lceil \log_d |\grp G| \rceil + \log_d  \alpha^{-1}$.

 \end{document}